\documentclass{aastex631}

\usepackage{amsmath}

\shortauthors{Faucher \& Blanton}

\graphicspath{{./}{figures/}}

\begin{document}

\title{A dust attenuation model inspired by the NIHAO-SKIRT-Catalog}

\author{Nicholas Faucher}
\affiliation{Center for Cosmology and Particle Physics, Department of Physics, New York University,
726 Broadway, New York, New York 10003, USA}

\author{Michael R. Blanton}
\affiliation{Center for Cosmology and Particle Physics, Department of Physics, New York University,
726 Broadway, New York, New York 10003, USA}

\begin{abstract}
\noindent We use simulated attenuation curves from the NIHAO-SKIRT-Catalog to test the flexibility of commonly used dust attenuation models in the face of the variations expected from realistic star-dust geometries. Motivated by lack of flexibility in these existing models, we propose a novel dust attenuation model with three free parameters that can accurately recover the simulated attenuation curves as well as the best-fitting curves from the commonly used models. This new model is fully analytic and treats all starlight equally, in contrast to two-component dust attenuation models. We use the parametrization
to investigate the relationship between the overall attenuation law shape
and the strength of the 2175 \AA\ bump. Our results indicate
variation in star-dust geometry leads these features to correlate tightly,
with grayer attenuation curves having weaker bumps.
\end{abstract}

\section{Introduction} 
\label{Introduction}

\noindent Untangling the history of galaxies over cosmic time allows us to 
better understand the physical process affecting the formation of stars and 
the growth of galaxies. One path to investigating this history is spectral energy 
distribution (SED) modeling of photometry or spectra of low redshift galaxies to infer 
their star formation and chemical evolution histories 
\citep{conroy10a, conroy13a}. An important uncertainty in SED modeling is 
dust, which reddens and attenuates the light in a manner that depends on
geometry, viewing angle, dust mass, and dust composition
\citep{hayward15a, trcka20a, lower20a, faucher23a}.  

This paper explores how well various commonly used dust attenuation models 
can reproduce the attenuation curves predicted from radiative transfer 
post-processing of high-resolution hydrodynamical galaxy simulations from 
the Numerical Investigation of Hundred Astrophysical Objects (NIHAO) 
project \citep{wang15a, faucher23a}. Our goals are to understand the potential
uncertainties in commonly used models and to develop techniques to reduce 
these uncertainties. Although our exploration here of dust attenuation models
is based on simulations run to redshift zero, astronomers apply the dust attenuation 
models we test at all redshifts, and therefore our results may be applicable 
to high redshift as well.

We expect that SED modeling inferences of star formation rates (SFRs) will be
sensitive to the assumed attenuation curve. For example, 
the slope of the ultraviolet SED yields information
about the attenuation if the attenuation curve is steep, but might not if it
is relatively gray; assuming the wrong attenuation curve models will yield
wrong total attenuation inferences and therefore wrong SFRs 
\citep{conroy10c}. Similar issues 
can arise in the analysis of optical photometry and spectra, affecting not
only SFRs but also stellar masses and other aspects of the star formation
history. \citep{greener20a}. \cite{salim20a} show that the inferred physical parameters of simulated galaxies can vary dramatically when changing the assumed dust attenuation curve. 

Since attenuation curves are not directly observable, many models have been proposed to explain the observed photometry of galaxies in the local Universe. In this work, we focus on the dust attenuation models of \cite{calzetti00a}, \cite{charlot00a}, \cite{kriek13a}, and \cite{cardelli89a}, which we refer to as Calzetti, Power Law, Kriek and Conroy, and Cardelli, respectively. Of these four dust attenuation models, only the Calzetti and Power Law attenuation models do not contain a UV-bump centered around 2175 \AA. We also note that all of these models except for Calzetti are two-component models, in which the light from young stellar populations receives additional extinction from a power law curve with its own free parameters.  

To improve our understanding of the accuracy of SED modeling inferences, in this
paper we test the flexibility of existing dust attenuation models by performing 
fits to attenuation curves in mock observations from the 
NIHAO-SKIRT-Catalog \citep{faucher23a}, which  take into consideration the 
three-dimensional distribution of stars and dust of NIHAO simulated galaxies, 
realistic dust grain size distributions and chemical compositions following 
The Heterogeneous dust Evolution Model for Interstellar Solids (THEMIS) dust 
model \citep{jones17a}, and scattering by dust into and out of the line 
of sight using the Stellar Kinematics Including Radiative Transfer 
(SKIRT) \citep{camps20a} Monte Carlo radiative transfer code.

Section \ref{Methods} describes how the simulated attenuation curves were 
calculated, the models that we compare to, and our fitting procedures. 
Section \ref{Results} presents the results. Section \ref{Discussion} discusses
the implications of the results, and we conclude in Section \ref{Conclusion}.

\section{Methods} 
\label{Methods}

\noindent In this section, we describe our sample of simulated attenuation curves,
the dust attenuation models we use, and our fitting procedure. We have released Python implementations of all of the dust attenuation models used in this work\footnote{\url{https://github.com/ntf229/dustModels}}.

\subsection{NIHAO-SKIRT-Catalog attenuation curves} \label{NSC}

\noindent In order to test the flexibility of dust attenuation models, we utilize mock photometry from the NIHAO-SKIRT-Catalog \citep{faucher23a} which includes the effects of realistic star-dust geometries from the NIHAO galaxy simulations \citep{wang15a}. In addition to performing full 3D radiative transfer, this mock catalog utilizes subgrid post-processing recipes that mitigate limitations in the temporal and spatial resolution of the simulations in order to produce far-ultraviolet (FUV) to far-infrared (FIR) photometry that statistically reproduce the colors of galaxies in the nearby Universe. The unattenuated stellar spectra in both the NIHAO-SKIRT-Catalog and those used in this work are generated from FSPS \citep{conroy09a, conroy10a, foreman-mackey14a} using the MIST isochrones \citep{paxton11a, paxton13a, paxton15a, choi16a, dotter16a}, the MILES spectral library \citep{sanchez06a}, and assuming a Chabrier stellar initial mass function (IMF) \citep{chabrier03a}.

\subsection{Calzetti} \label{Calzetti}

\noindent The Calzetti \citep{calzetti00a} dust extinction curve is the 
simplest model we explore in this 
work. In the most flexible case, it is controlled by only two parameters: 
The overall scaling, which we set by the attenuation in the V-band, 
$A_{\rm V} \in [0,10]$, and a variable fraction of unattenuated stellar light, 
$f_{\rm no-dust} \in [0,1]$, where the indicated ranges are those allowed in the 
fitting procedure. Allowing some of the stellar light to reach the observer 
unattenuated leads to more gray attenuation curves. This extinction law has 
been used in in many influential works, e.g. investigating the environment-dependent
star formation rate and mass evolution \citep{peng10a}, associating 
galaxy growth history with dark matter halo assembly history \citep{behroozi19a}, 
and estimating photometric redshifts for James Webb Space Telescope
catalogs \citep{castellano22a, treu22a, adams23a}.

\subsection{Power Law} \label{Power Law}

\noindent The Power Law extinction curve is a two-component model due to 
\citep{charlot00a}, meaning that the 
young (age$ \, < \, 10$ Myr) stellar populations are attenuated by an additional 
power law extinction curve with its own free parameters, before being added to 
and attenuated along with the rest of the stellar populations. These two components 
are meant to represent the attenuation from molecular clouds surrounding young 
stellar populations and the attenuation from the diffuse interstellar medium (ISM). 
This attenuation model has 6 free parameters: The overall scaling of the diffuse 
attenuation, $A_{\rm V} \in [0,10]$, the power law index of the diffuse attenuation, 
$p \in [-5,0]$, the overall scaling of the young attenuation, $A_{\rm V, \, young} 
\in [0,10]$, the power law index of the young attenuation, $p_{\rm young} \in 
[-5,0]$, and a variable fraction of unattenuated stellar light for both the 
diffuse and young attenuation, $f_{\rm no-dust} \in [0,1]$, and $f_{\rm no-dust, \, 
young} \in [0,1]$, respectively. This attenuation law was assumed in the heavily
used MPA-JHU catalog (\citealt{kauffmann03a, tremonti04a, brinchmann04}) including 
studies of AGN host galaxies, the origin of the mass-metallicity relation,  
the mass dependence of the star formation rates of galaxies, and many others.

\subsection{Kriek and Conroy} \label{Kriek and Conroy}

\noindent The Kriek and Conroy \citep{kriek13a} extinction curve is also a two-component model.
The young stellar populations are attenuated by a power law extinction 
curve with its own free parameters. Then, an additional layer of ``diffuse'' attenuation 
is applied uniformly to the entire stellar population. This model has 6 
free parameters: The overall scaling of the diffuse attenuation, 
$A_{\rm V} \in [0,10]$, the power law index of the diffuse attenuation
$p \in [-5,0]$, 
the overall scaling of the young attenuation, $A_{\rm V, \, young} \in [0,10]$, 
the power law index of the young attenuation, $p_{\rm young} \in [-5,0]$, and a 
variable fraction of unattenuated stellar light for both the diffuse and 
young attenuation, $f_{\rm no-dust} \in [0,1]$, and $f_{\rm no-dust, \, young} \in [0,1]$, respectively. The diffuse attenuation has a 2175 \AA \,UV bump with a strength set deterministically by the power law slope, $p$, with steeper slopes corresponding to stronger UV-bumps. However, the effective power law slope of this model is not given by $p$, but rather $\sim (p - 0.8)$ due to the power law term being multiplied by the Calzetti curve, which also causes the resulting attenuation curves to deviate from the power law shape even when excluding the UV bump term:

\begin{equation}
A_{\lambda} =  \, \frac{A_V}{4.05} \, \left(k \left( \lambda \right) + D \left( \lambda \right) \right) 
 \left(\frac{\lambda}{5500 \, \mathrm{\AA}}\right)^{p}  
\label{eq:KC}
\end{equation}
where $k\left(\lambda \right)$ is the Calzetti cruve, $\lambda$ has units of \AA,  
\begin{equation}
D \left( \lambda \right) \, = \, \frac{E_{b} \left(\lambda \, \Delta \lambda \right)^2}{\left( \lambda^2 - \lambda_{0}^2 \right)^2 +  \left(\lambda \, \Delta \lambda \right)^2} 
\label{eq:KC_bump}
\end{equation}
with $E_{b} \, = \, 0.85 - 1.9 \, p $, $\Delta \lambda = 350 \, \mathrm{\AA}$, and $\lambda_{0} = 2175 \, \mathrm{\AA}$.

This model has been used to investigate the progenitors of ultra-massive 
galaxies \citep{marchesini14a}, the evolution of the mass-size relation 
\citep{suess19a}, and the growth of metals and stellar mass at high 
redshift \citep{tecchella22a}. 

\subsection{Cardelli} \label{Cardelli}

\noindent The Cardelli \citep{cardelli89a} extinction curve, also known as the 
Milky Way extinction or CCM (Cardelli, Clayton, and Mathis) law, 
is also a two-component model of the same type as above. In this case the 
young stellar population attenuation is again a power law, and the diffuse
attenuation is a Milky Way extinction curve. In contrast to the original model used in \cite{cardelli89a}, we use the model as incorporated in FSPS which includes a variable UV-bump term. The model has 7 free parameters: The overall scaling of the diffuse attenuation, 
$A_{\rm V} \in [0,10]$, the ratio of total to selective absorption which characterizes 
the Milky Way extinction curve, $R_{\rm V} \in [0,10]$, 
the strength of the 2175 \AA \,UV-bump, $b_{\rm UV} \in [0,10]$, the 
overall scaling of the young attenuation, $A_{\rm V, \, young} \in [0,10]$, the power 
law index of the young attenuation, $p_{\rm young} \in [-5,0]$, and a variable fraction 
of unattenuated stellar light for both the diffuse and young attenuation, $f_{\rm no-
dust} \in [0,1]$, and $f_{\rm no-dust, \, young} \in [0,1]$, respectively. Although this model has been widely used to account for the foreground extinction curve of light by the Milky Way's ISM in the context of extragalactic surveys (\citealt{schlegel98a} 
among literally thousands
of others), it has also been used to account for the dust attenuation of host galaxies in SED modeling such as in the study of the progenitors of ultra-massive galaxies \citep{marchesini14a} and in order to find correlations between galactic spectral type and the best-fitting dust attenuation law \citep{kriek13a}.

\subsection{TEA Model} \label{TEA Model}

\begin{figure}
\begin{center}
\includegraphics[width=0.6\textwidth]{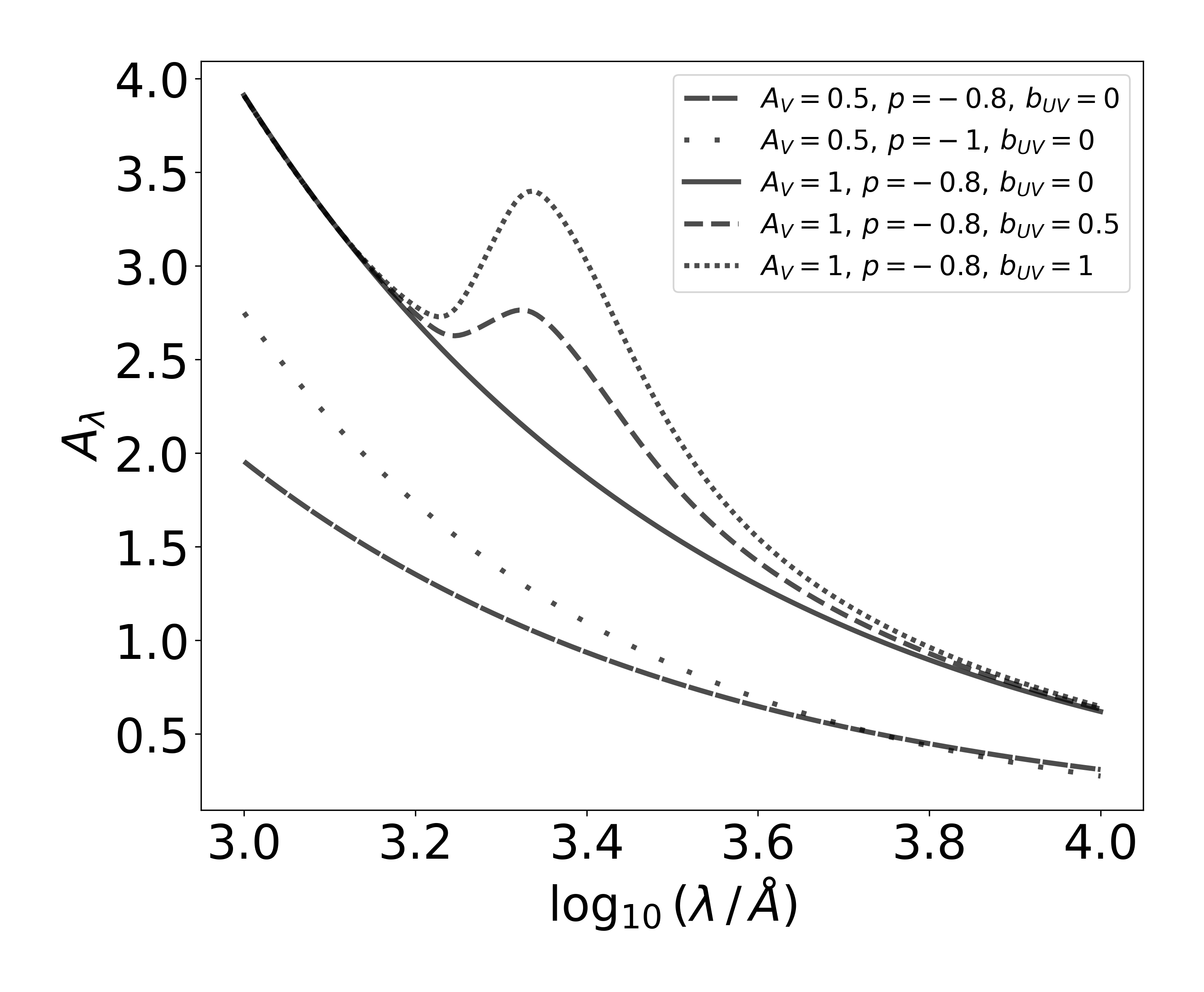}
\end{center}
\caption{\label{fig:plotTEA_varyParams} \small Visualization of TEA model attenuation curves in magnitudes over a range of parameters.}
\end{figure}

\noindent Our proposed model, which we call the THEMIS Effective Attenuation 
(TEA) model, has the following functional form with 3 free parameters 
$A_{\rm V} \in [0,10]$, $p \in [-5,0]$, and $b_{\rm UV} \in [0,10]$:

\begin{equation}
A_{\lambda} =  \, A_{\rm V} \, \left(\frac{\lambda}{5500 \, \mathrm{\AA}}\right)^{p}  
+ \,  b_{\rm UV} \left\{ \left[1+{\rm erf}\left( s_{\rm UV} \, f_{1}\left(\lambda\right) \right) \right] \left(1+\left(6 f_{1}(\lambda)\right)^{2} \right)^{-3/2}\, -  \frac{1}{4} \left( \frac{ f_{2}(\lambda) ^{5}} 
{f_{2}(\lambda) ^{4} + 1} + f_{2}(\lambda) \right) \right\} \label{eq:tea}
\end{equation}
where $\lambda$ has units of \AA, $s_{UV}=6.95$,  
\begin{eqnarray}
f_{1}(\lambda) &=& \log_{10}(\lambda)-3.312,\cr
f_{2}(\lambda) &=& 16 \left( \log_{10}(\lambda)-3.285 \right), 
\end{eqnarray}
and ${\rm erf}()$ is the error function. The first term is a power law with variable strength 
($A_{\rm V}$) and power-law slope ($p$). The second term is a skewed bump function where the bump 
strength is controlled by the parameter $b_{\rm UV}$ and the skewness is set 
by $s_{\rm UV}$. The last term scales with the bump strength $b_{\rm UV}$ 
and adds attenuation to the bump while subtracting attenuation from wavelengths 
slightly shorter than the bump. We can see from Figure \ref{fig:plotTEA_varyParams}, which shows attenuation curves from the TEA model over a range of parameters, that the value of $A_{\rm V}$ only corresponds to the actual attenuation in the V-band ($\log_{10}(\lambda_{\rm eff} \, / \, \mathrm{\AA})=3.74$) when $b_{\rm UV}=0$, since $A_{\rm V}$ is the attenuation in the V-band for the power law component only and the UV-bump extends to large wavelengths.  

Equation \eqref{eq:tea} is not a dust extinction curve, it is a dust attenuation curve 
that was designed to encapsulate the effects of the star-dust geometry from the 
NIHAO simulated 
galaxies and the subgrid attenuation from star-forming regions as modeled by the 
photo-ionization code MAPPINGS-III using Starburst99 SEDs \citep{leitherer99a} as 
implemented in SKIRT.

\subsection{Fitting Procedure} \label{Fitting Procedure}

\noindent We test the flexibility of these dust attenuation models by fitting them to attenuation curves from the NIHAO-SKIRT-Catalog. We fit each dust model to these simulated attenuation curves by minimizing the root mean squared error (RMSE) of $A_{\lambda}$ in magnitudes on a logarithmically spaced grid of 75 wavelengths between $10^3$ and $10^4$ \AA \, using the dual annealing global minimization function \citep{xiang97a} implemented in {\tt scipy.optimize} \citep{virtanen20a}, set to 2000 maximum iterations. 
In our fits, we will test both with and without freedom in $f_{\rm no-dust}$ and 
$f_{\rm no-dust, \, young}$. We also note that we are allowing more flexibility in these dust models than is typically done in SED modeling. For instance, both \cite{leja17a} and \cite{hahn22a} implement the Kriek and Conroy dust attenuation model within the SED modeling frameworks Prospector and SEDFLOW, respectively, but only allow the $A_{\rm V}$, $p$, and $A_{\rm V, \, young}$ attenuation parameters to vary.  

The two-component dust models require a model for the star formation history to 
define an attenuation curve. Here we use the most favorable choice to these models,
which is the actual star formation history in these NIHAO galaxies. 
For this star formation history we use the chemical evolution histories (CEHs) 
published in the 
NIHAO-SKIRT-Catalog, which give the star formation rate as a function of age and
metallicity.
We separate the populations into those greater and less than 10 Myr in age. Then we apply each dust attenuation model component to the dust-free spectra of 
the young and old populations, and construct an attenuated spectrum $f_A(\lambda)$. The 
attenuation curve is then $-2.5\log_{10} f_A(\lambda) / f(\lambda)$, where $f(\lambda)$
is the unattenuated spectrum.

\section{Results} 
\label{Results}

\noindent In Section \ref{Fitting attenuation models to the NIHAO-SKIRT-Catalog}, we explore the ability of each dust model to recover the attenuation curves from dusty galaxies from the NIHAO-SKIRT-Catalog. In Section \ref{Fitting the TEA model to best-fitting common model attenuation curves}, we see how well the newly proposed TEA model can recover the best fits of the other dust models. Finally, we use the NIHAO-SKIRT-Catalog fits from the TEA model to explore correlations between its parameters in Section \ref{Correlation Between TEA model parameters}. 

\subsection{Fitting attenuation models to the NIHAO-SKIRT-Catalog} \label{Fitting attenuation models to the NIHAO-SKIRT-Catalog}

\begin{figure}
\gridline{\fig{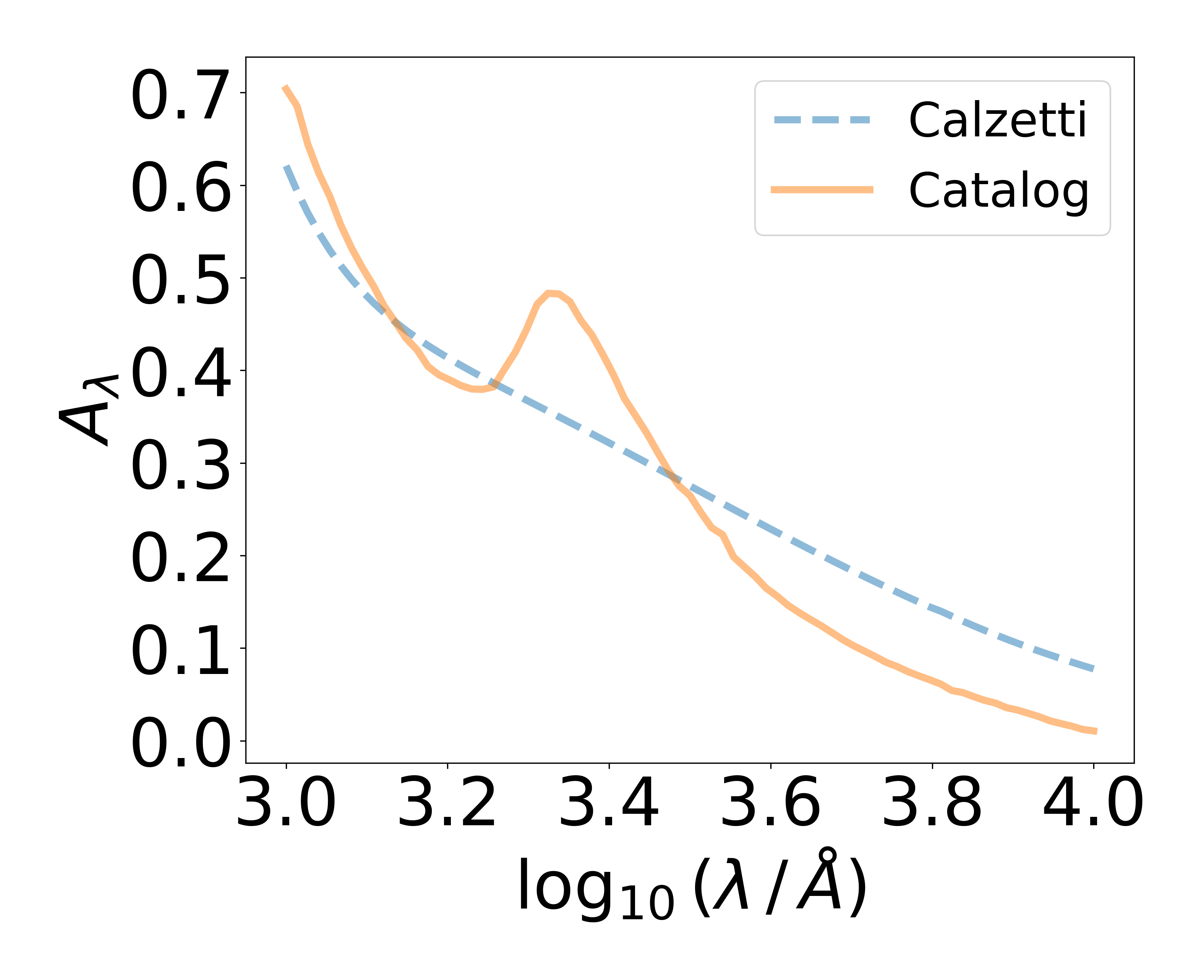}{0.3\textwidth}{(a) Calzetti}
          \fig{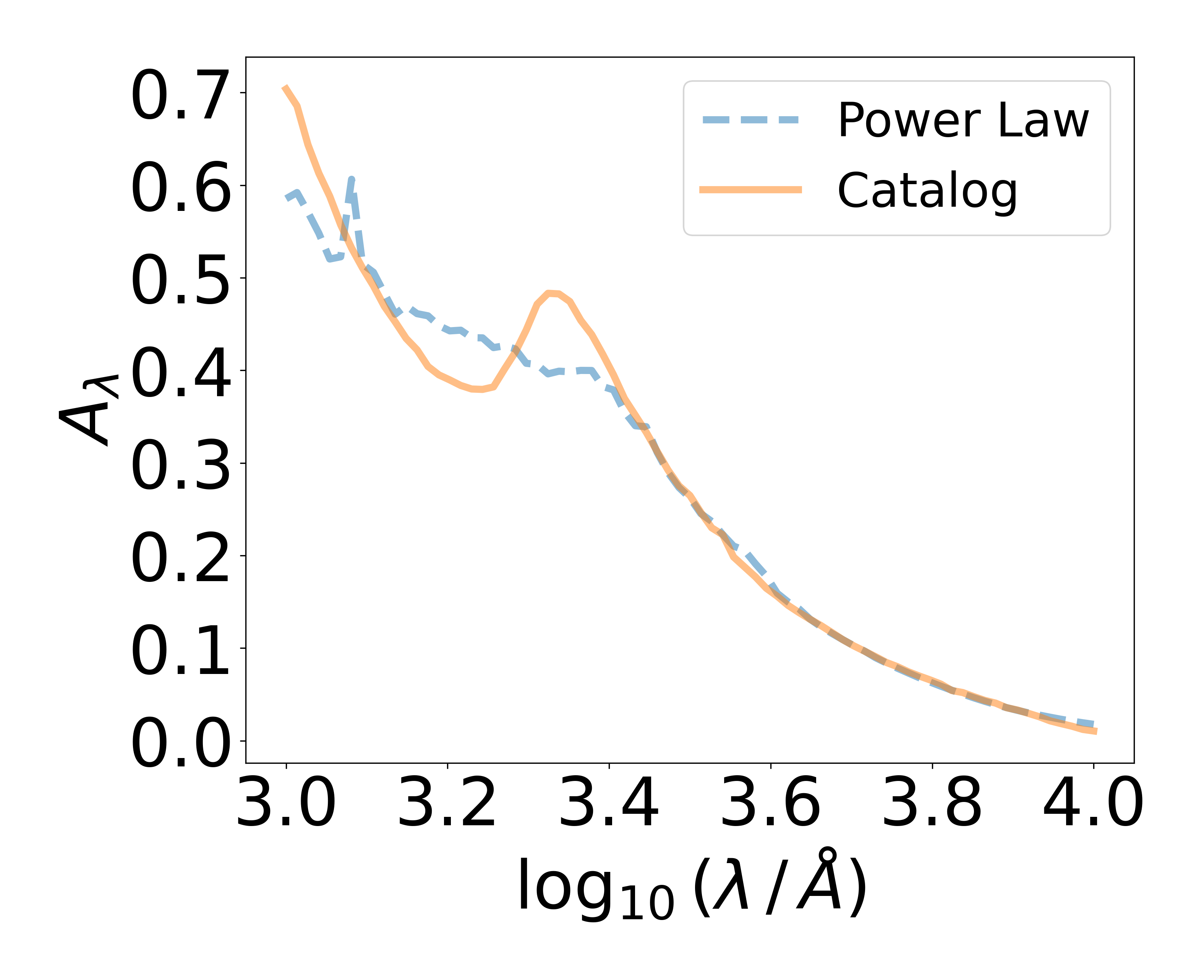}{0.3\textwidth}{(b) Power Law}
          \fig{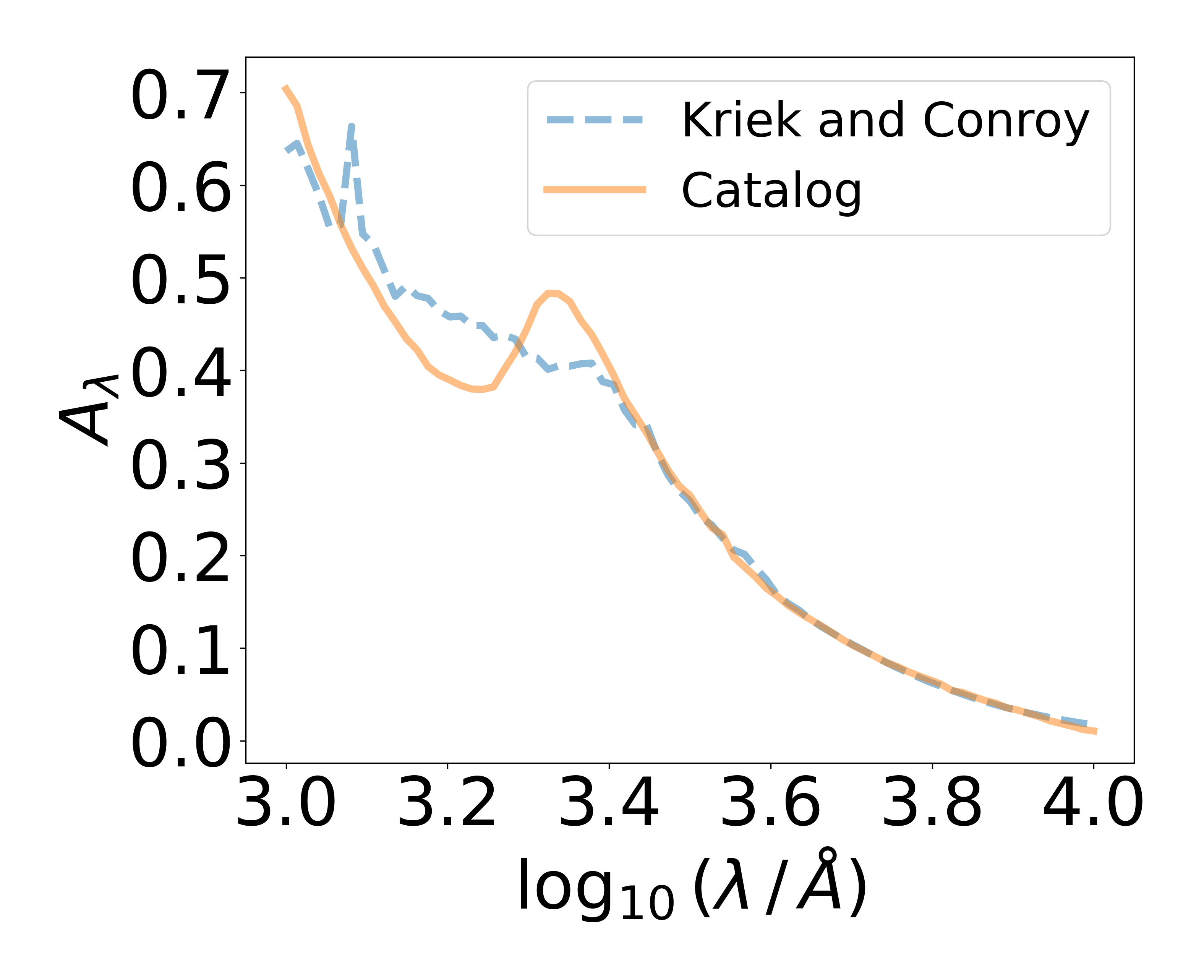}{0.3\textwidth}{(c) Kriek and Conroy}}
\gridline{\fig{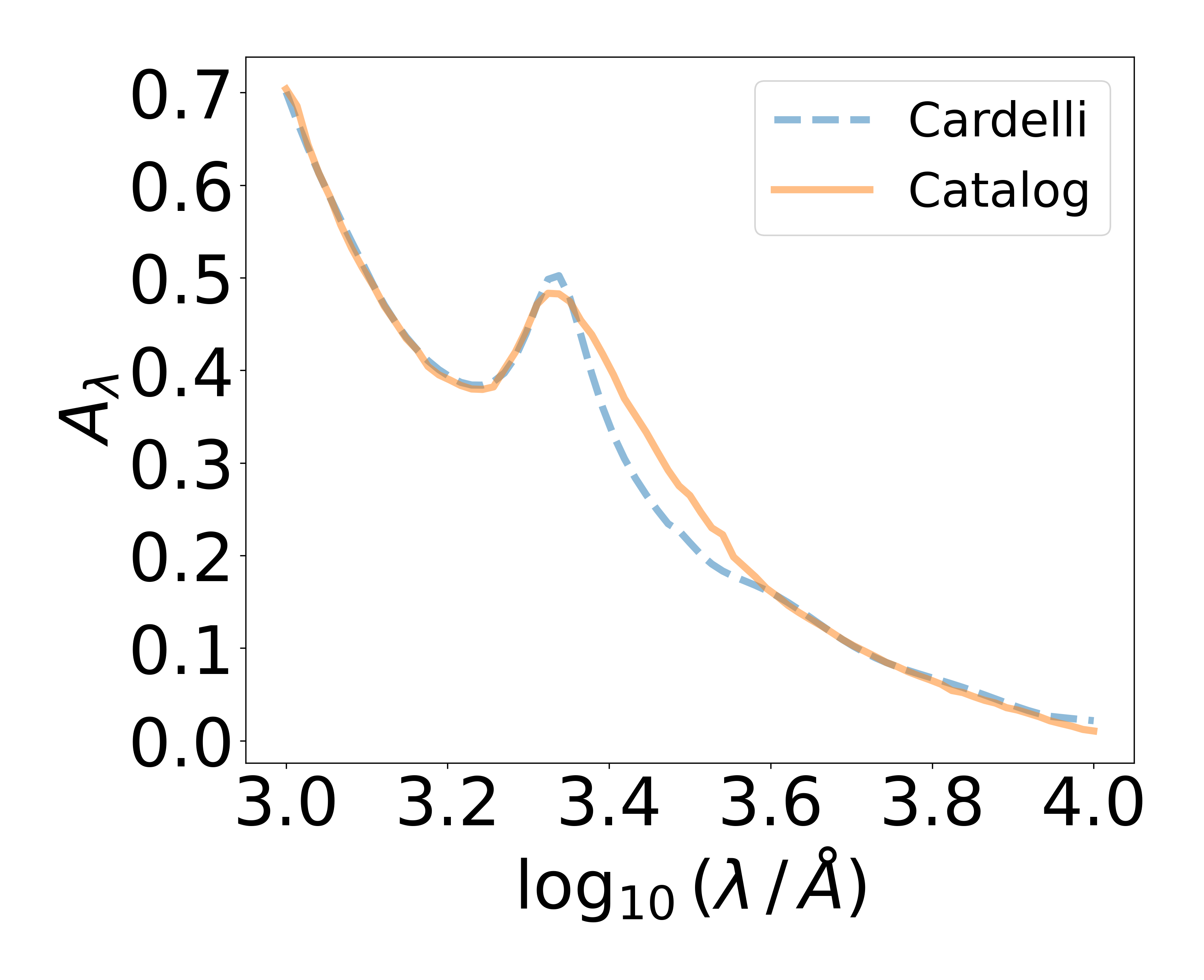}{0.3\textwidth}{(d) Cardelli}
          \fig{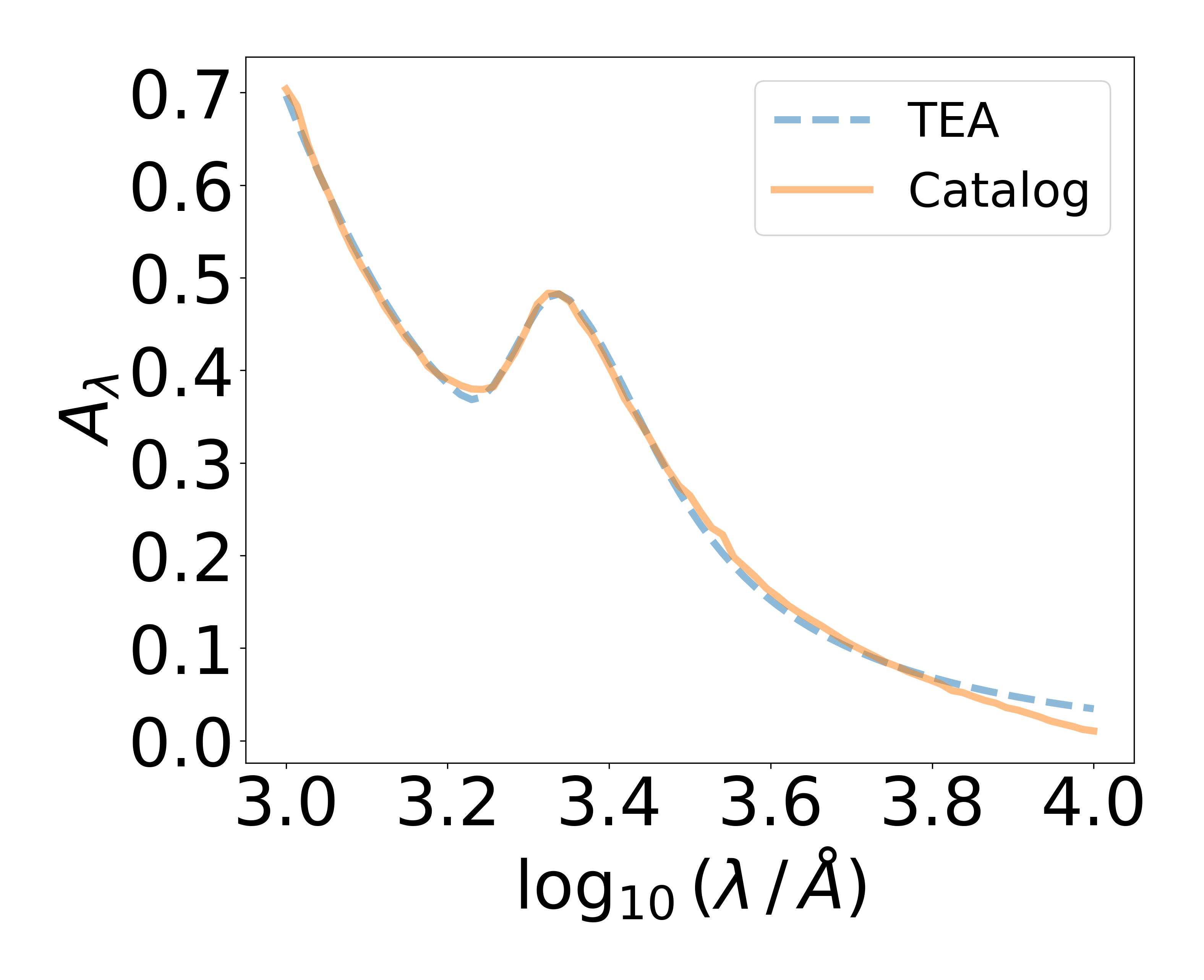}{0.3\textwidth}{(e) TEA}}
\caption{\label{fig:models1} Fits of dust attenuation models to face-on orientation of the attenuation curve in magnitudes from the simulated galaxy {\tt g5.02e11}, allowing all parameters to be free in the fits. In this case, Calzetti has 2 free parameters, Power Law has 6 free parameters, Kriek and Conroy has 6 free parameters, Cardelli has 7 free parameters, and the TEA model has 3 free parameters. }
\end{figure}

\noindent Figure \ref{fig:models1} shows the fits of the face-on orientation of the simulated galaxy {\tt g5.02e11} for the the five dust models being considered in this work. In this case, we allow all parameters to be free. In total, the Calzetti attenuation model has 2 free parameters ($A_{\rm V}$ and $f_{\rm no-dust}$), the Power Law attenuation model has 6 free parameters ($A_{\rm V}$, $p$, $A_{\rm V, \, young}$, $p_{\rm young}$, $f_{\rm no-dust}$, and $f_{\rm no-dust, \, young}$), the Kriek and Conroy attenuation model has 6 free parameters ($A_{\rm V}$, $p$, $A_{\rm V, \, young}$, $p_{\rm young}$, $f_{\rm no-dust}$, and $f_{\rm no-dust, \, young}$), the Cardelli attenuation model has 7 free parameters ($A_{\rm V}$, $R_{\rm V}$, $b_{\rm UV}$, $A_{\rm V, \, young}$, $p_{\rm young}$, $f_{\rm no-dust}$, and $f_{\rm no-dust, \, young}$), and the TEA model has 3 free parameters ($A_{\rm V}$, $p$, and $b_{\rm UV}$).

\begin{figure}
\begin{center}
\plottwo{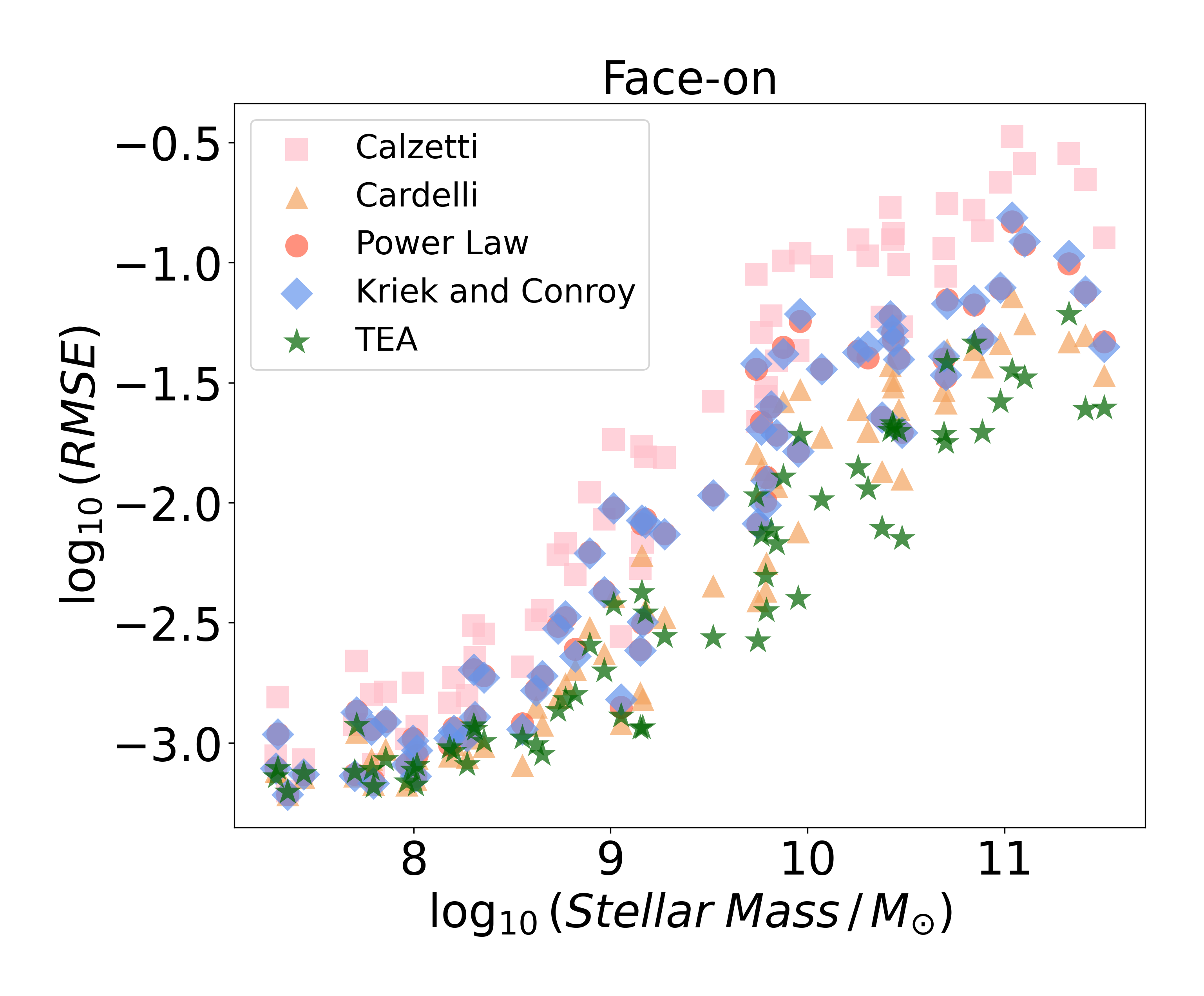}{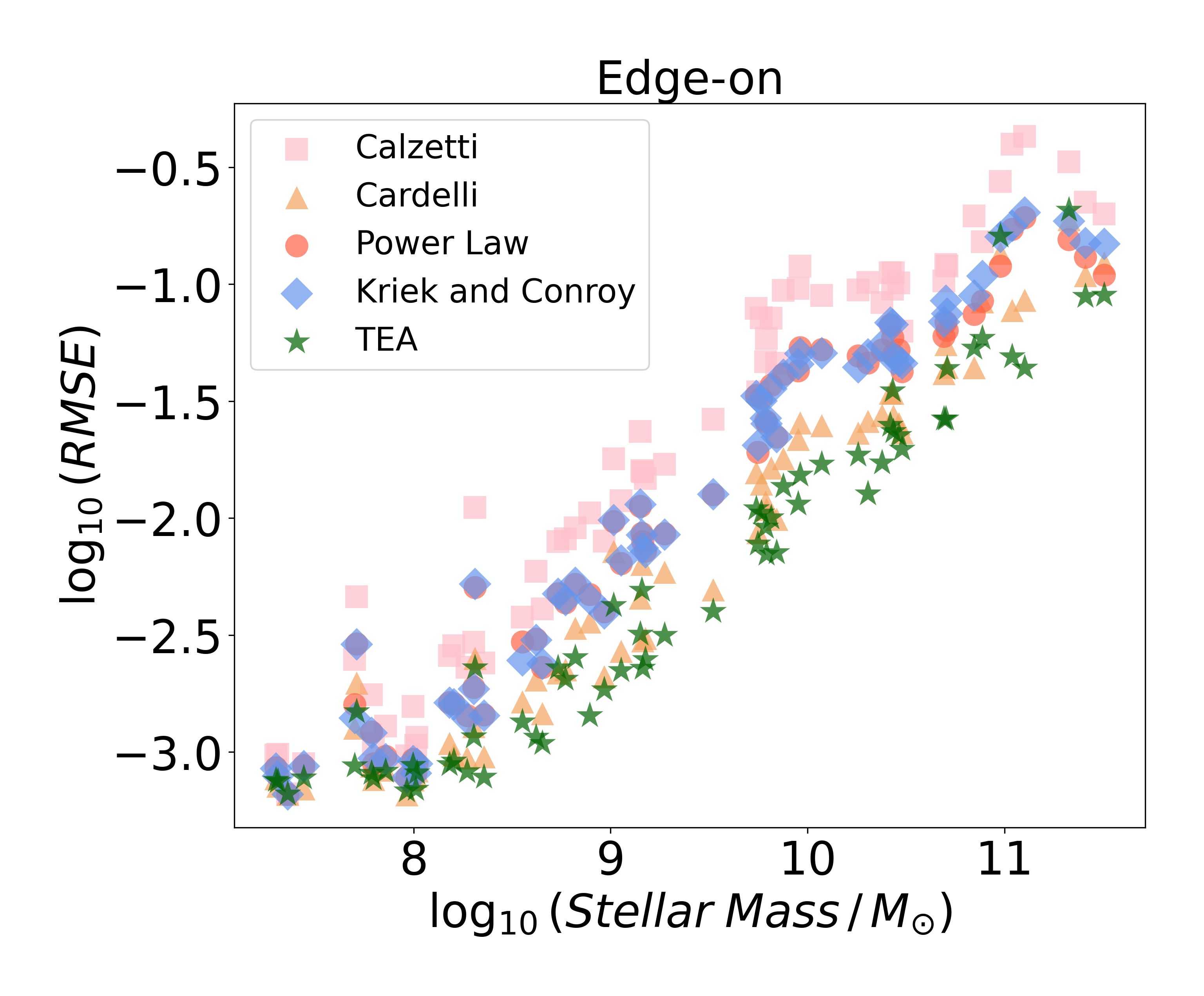}
\end{center}
\caption{\label{fig:errors1} \small 
Root mean squared errors (RMSE) of the dust attenuation model fits for the face-on (left) and edge-on (right) orientations of the simulated galaxies' attenuation curves in magnitudes, allowing all parameters to be free.}
\end{figure}

\begin{figure}
\begin{center}
\plottwo{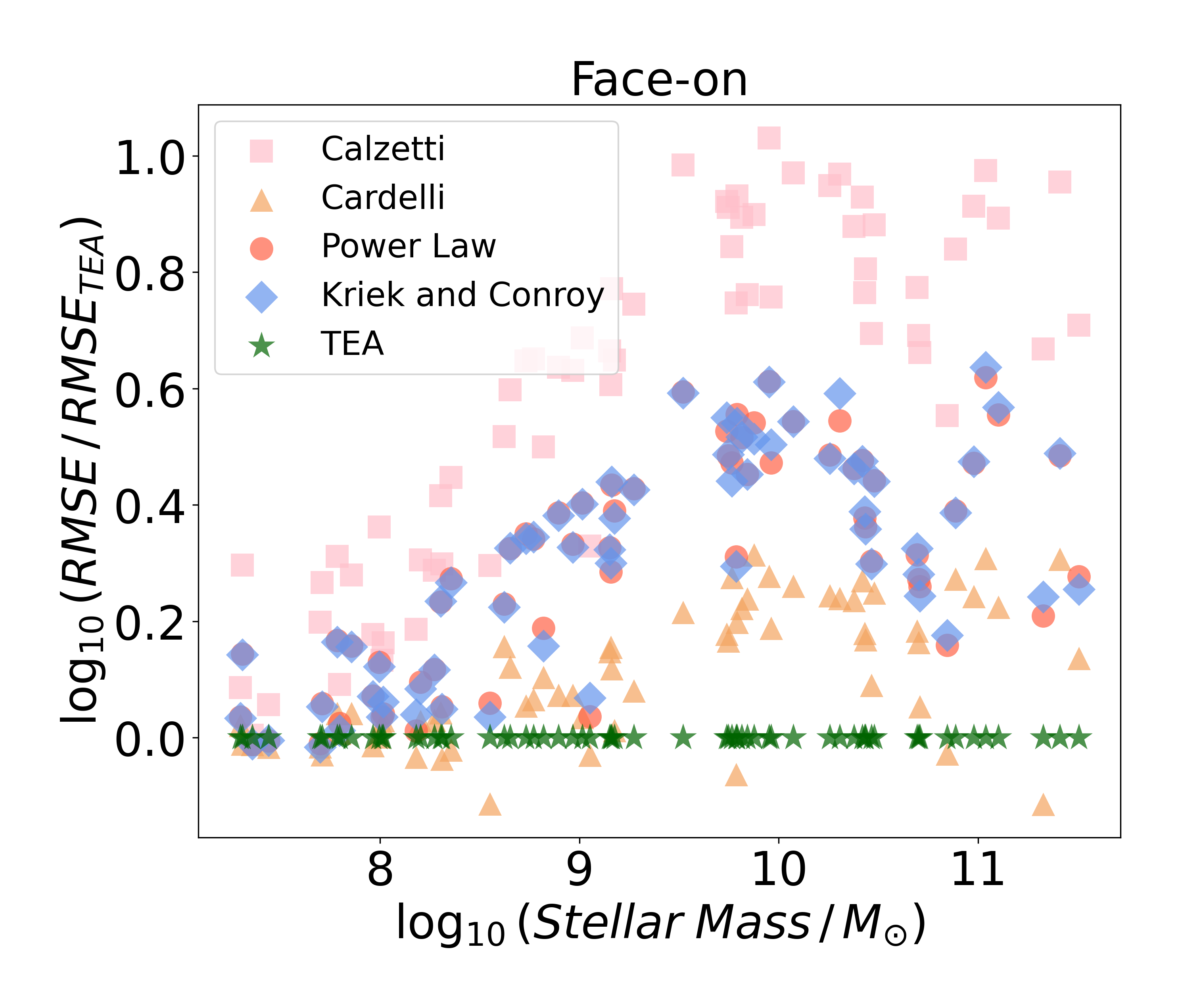}{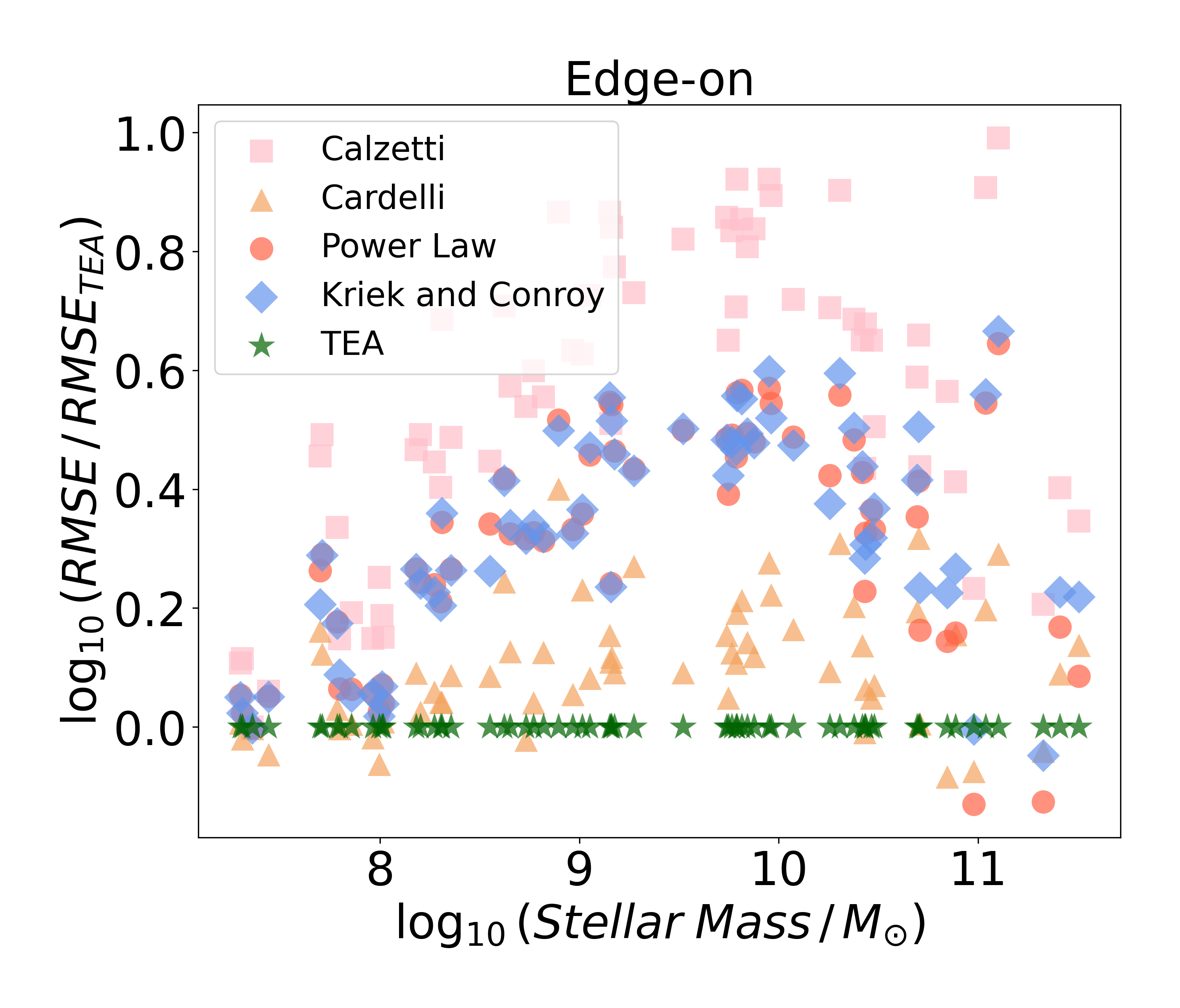}
\end{center}
\caption{\label{fig:shifted_errors1} \small 
Similar to Figure \ref{fig:errors1}, now with the RMSEs shifted by the TEA model's RMSE for each galaxy.}
\end{figure}

Figure \ref{fig:errors1} shows the root mean squared error (RMSE) in magnitudes as a function of galaxy 
stellar mass between the dust model fits and the simulated attenuation curves. 
In these fits we treat as free parameters $f_{\rm no-dust}$ and 
$f_{\rm no-dust, \, young}$, representing the fractions
of starlight allowed to reach the observer unattenuated. The TEA model 
outperforms all other models for nearly all galaxies and orientations. Note that these RMSEs are calculated from the unnormalized attenuation curves, which causes the general trend of increasing RMSEs with increasing stellar mass, since more massive galaxies tend to have stronger dust attenuation. Figure \ref{fig:shifted_errors1} shows the RMSE, offset in log-space by the RMSE of the TEA model, as a function of galaxy stellar mass. This plot makes it easier to see how the different models' RMSEs compare to each other by subtracting away the scatter from the differences in RMSEs between galaxies.

\begin{figure}
\gridline{\fig{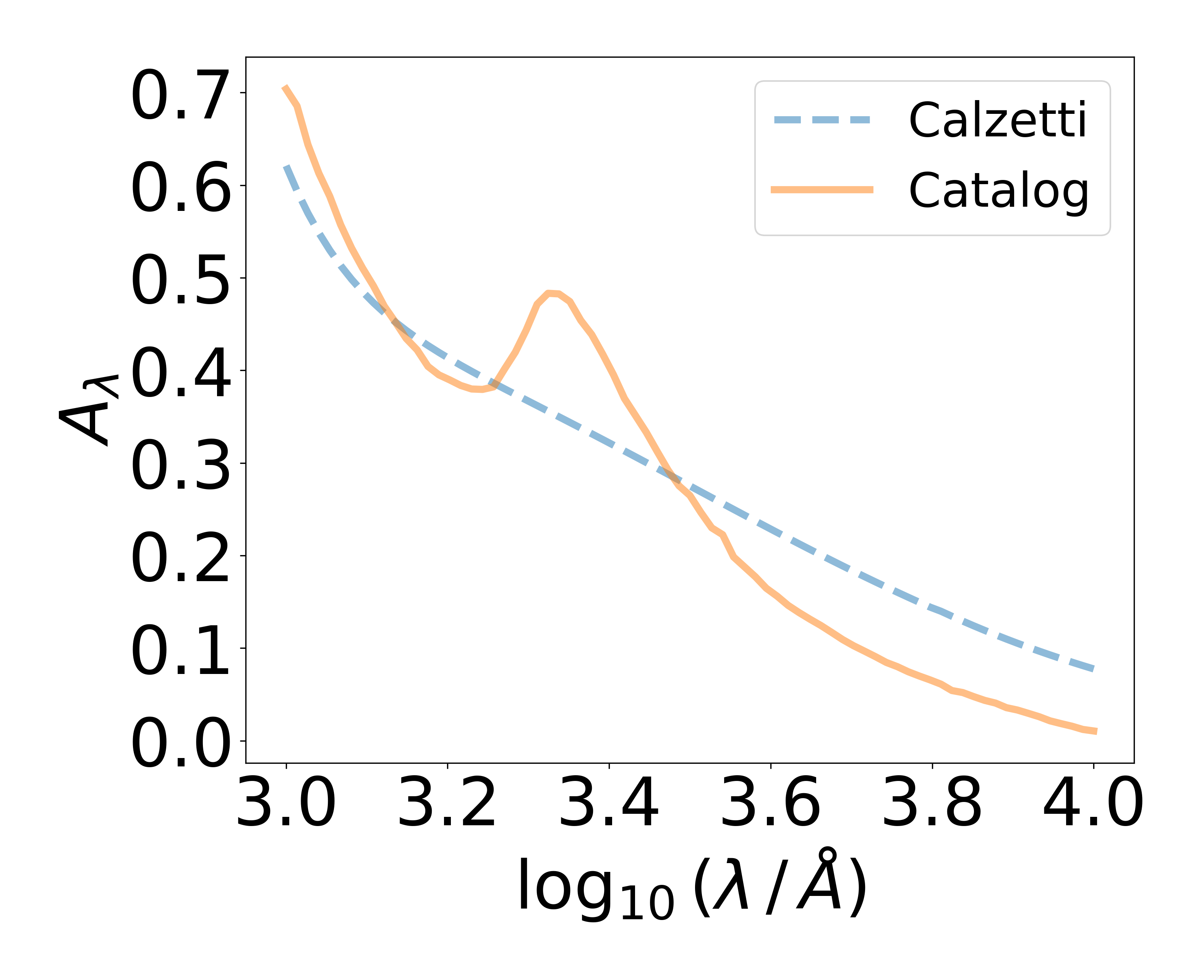}{0.3\textwidth}{(a) Calzetti}
          \fig{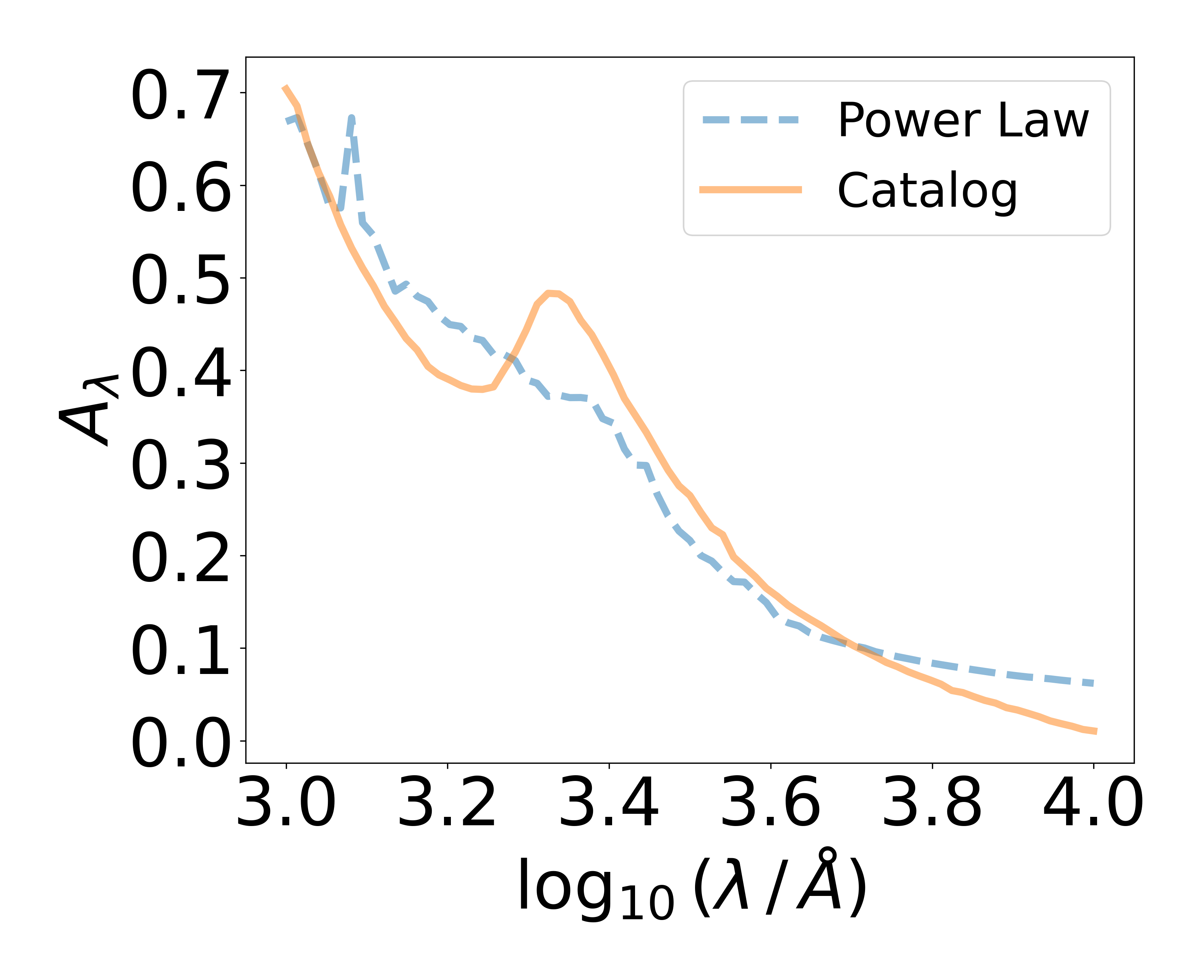}{0.3\textwidth}{(b) Power Law}
          \fig{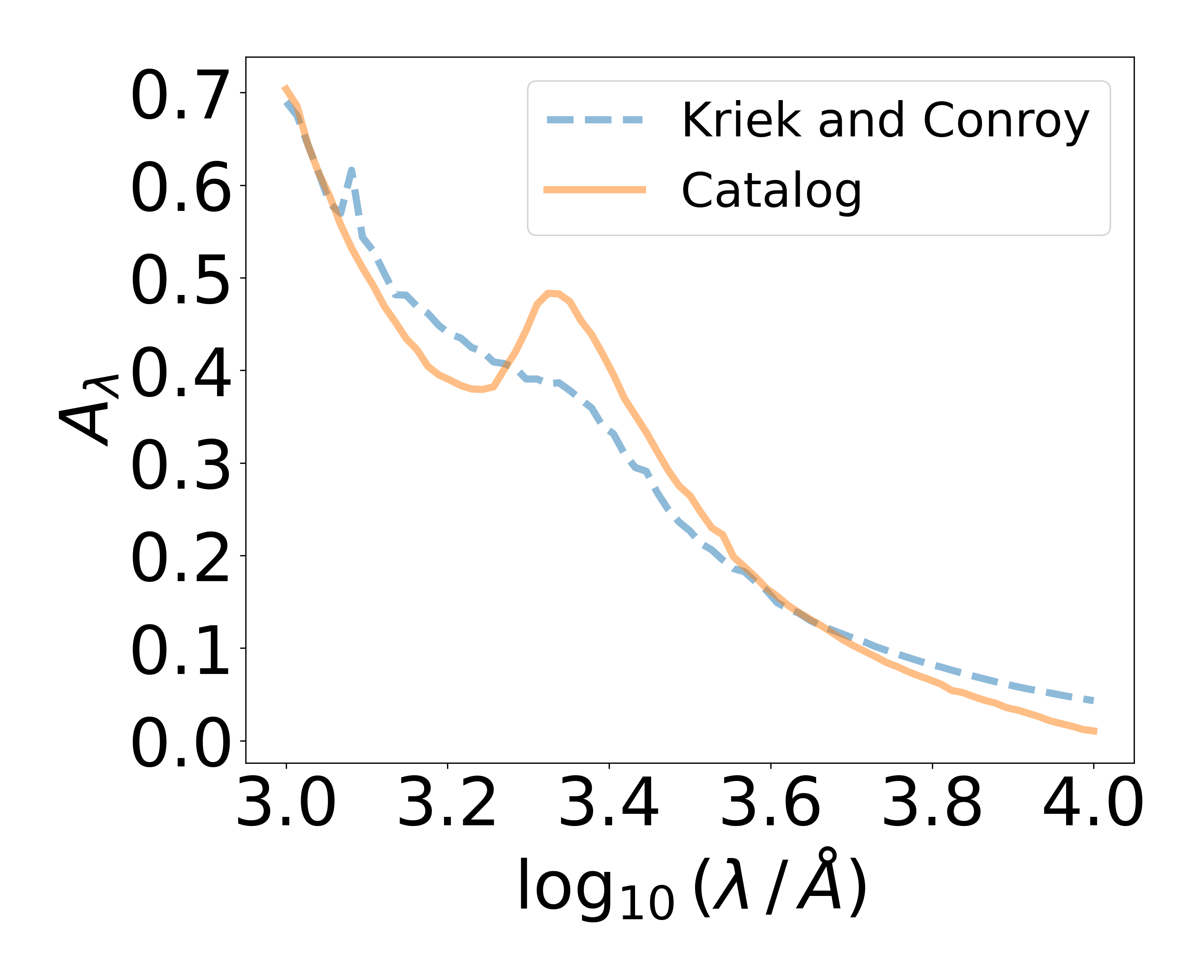}{0.3\textwidth}{(c) Kriek and Conroy}}
\gridline{\fig{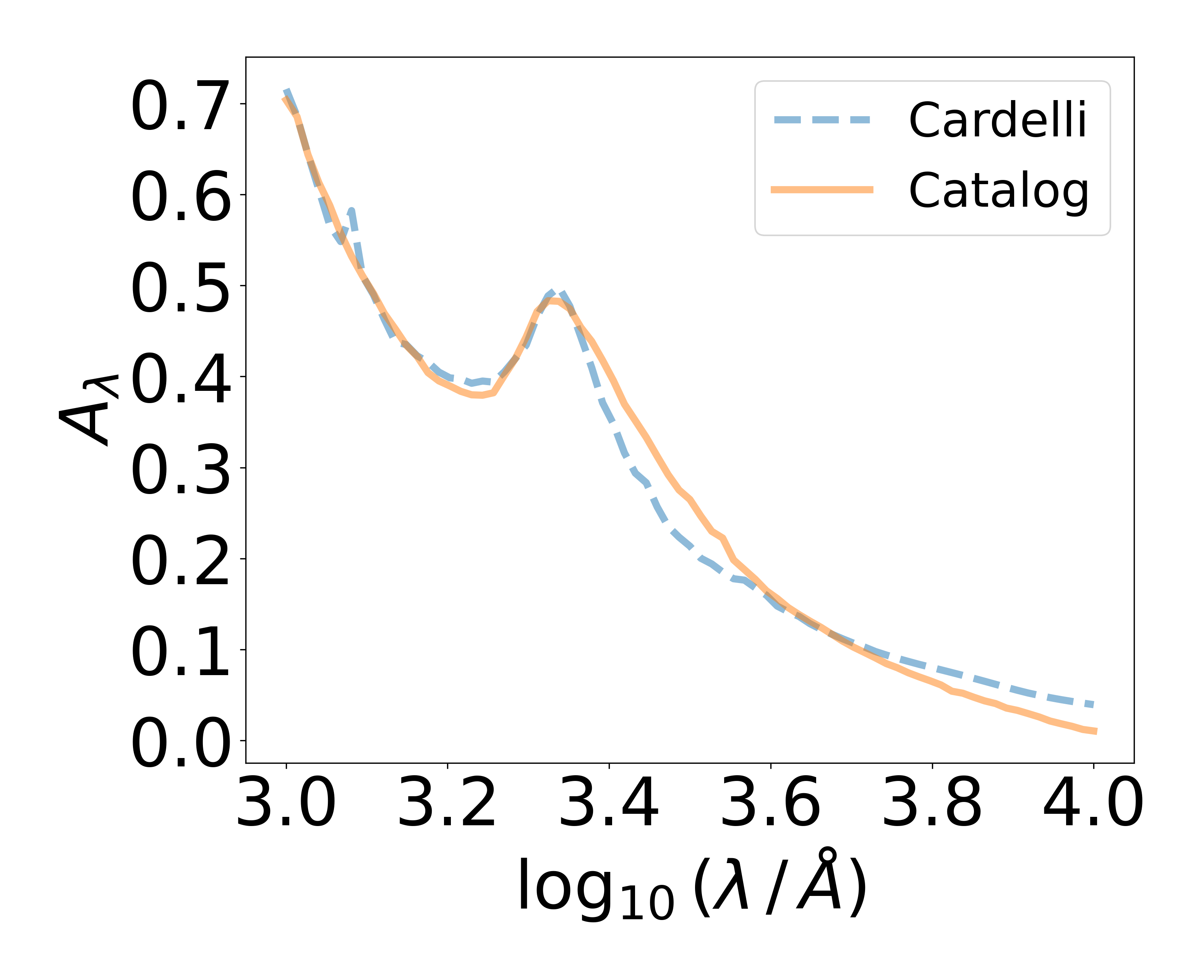}{0.3\textwidth}{(d) Cardelli}
          \fig{TEA_g5.02e11_face-on_attenuation.png}{0.3\textwidth}{(e) TEA}}
\caption{\label{fig:models2} 
Similar to Figure \ref{fig:models1}, now not allowing any fraction of the stellar light to reach the observer unattenuated.
In this case, Calzetti has 1 free parameter, Power Law has 4 free parameters, Kriek and Conroy has 4 free parameters, Cardelli has 5 free parameters, and the TEA model has 3 free parameters.}
\end{figure}

Figure \ref{fig:models2} shows the same fits as in Figure \ref{fig:models1}, but with the parameters $f_{\rm no-dust}$ and $f_{\rm no-dust, \, young}$ fixed to zero, meaning all starlight gets attenuated. In total, the Calzetti attenuation model has 1 free parameter ($A_{\rm V}$), the Power Law attenuation model has 4 free parameters ($A_{\rm V}$, $p$, $A_{\rm V, \, young}$, $p_{\rm young}$), the Kriek and Conroy attenuation model has 4 free parameters ($A_{\rm V}$, $p$, $A_{\rm V, \, young}$, $p_{\rm young}$), the Cardelli attenuation model has 5 free parameters ($A_{\rm V}$, $R_{\rm V}$, $b_{\rm UV}$, $A_{\rm V, \, young}$, $p_{\rm young}$), and the TEA model has 3 free parameters ($A_{\rm V}$, $p$, and $b_{\rm UV}$).

\begin{figure}
\begin{center}
\plottwo{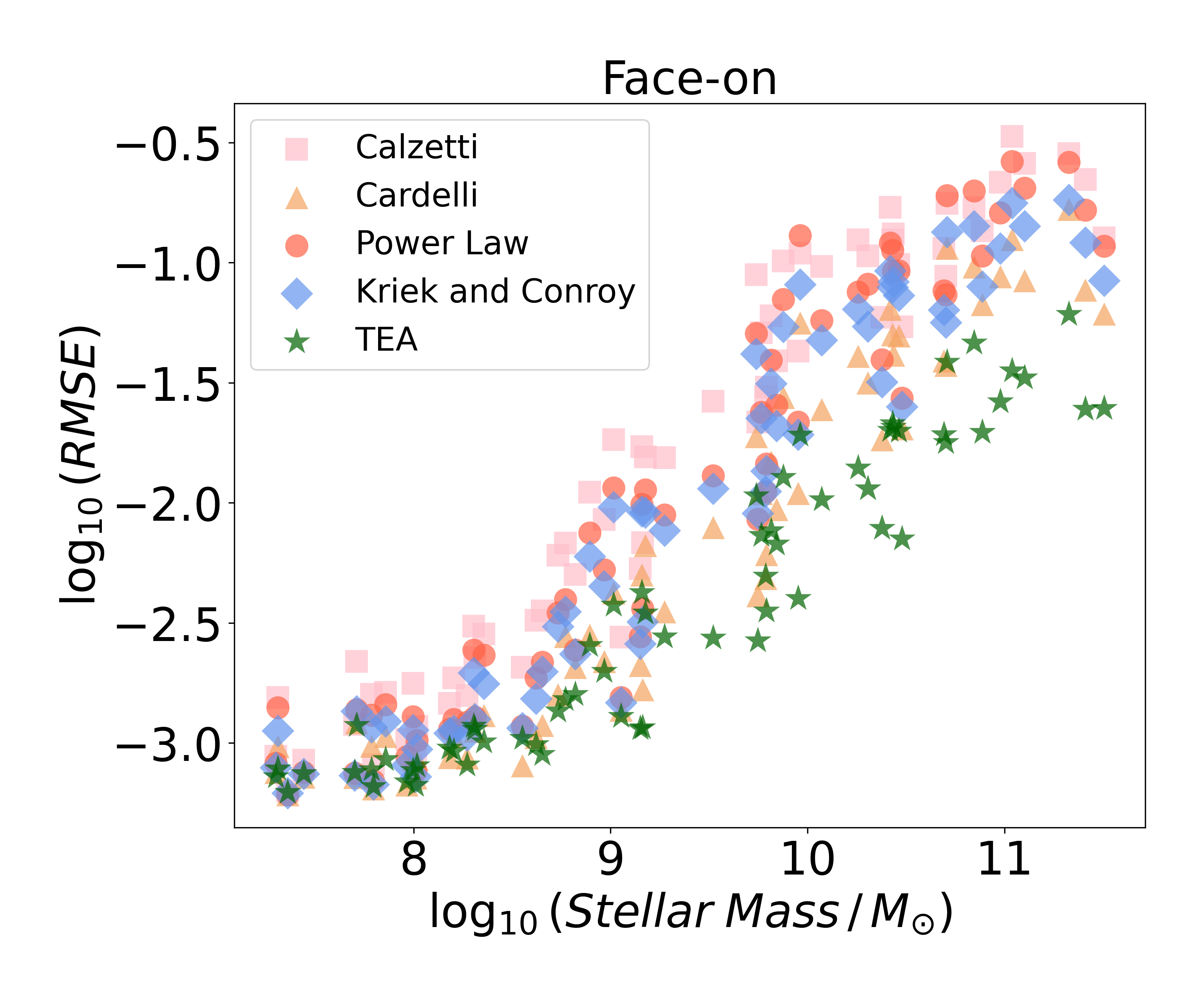}{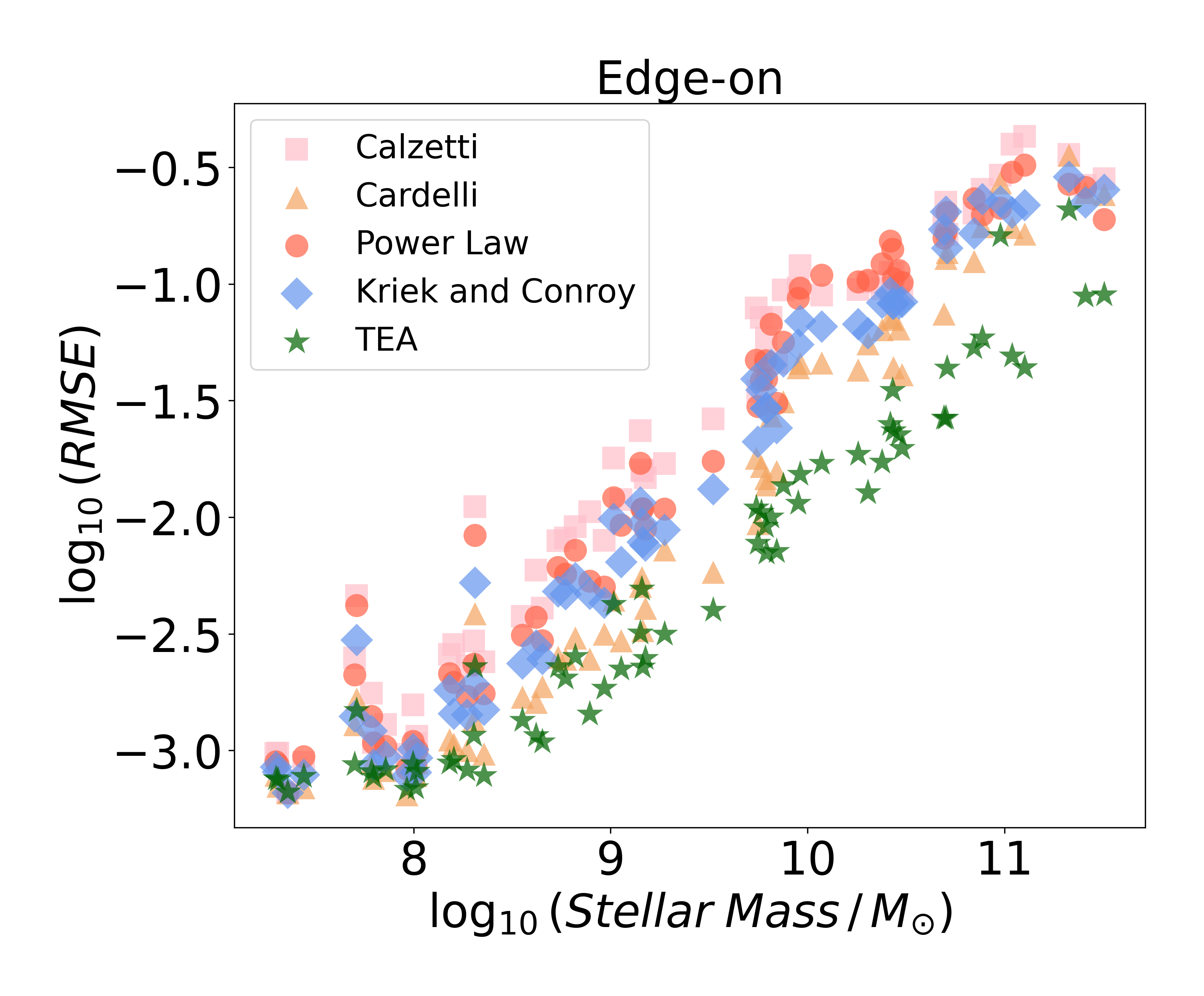}
\end{center}
\caption{\label{fig:errors2} \small 
Similar to Figure \ref{fig:errors1}, now not allowing any fraction of the stellar light to reach the observer unattenuated.}
\end{figure}

\begin{figure}
\begin{center}
\plottwo{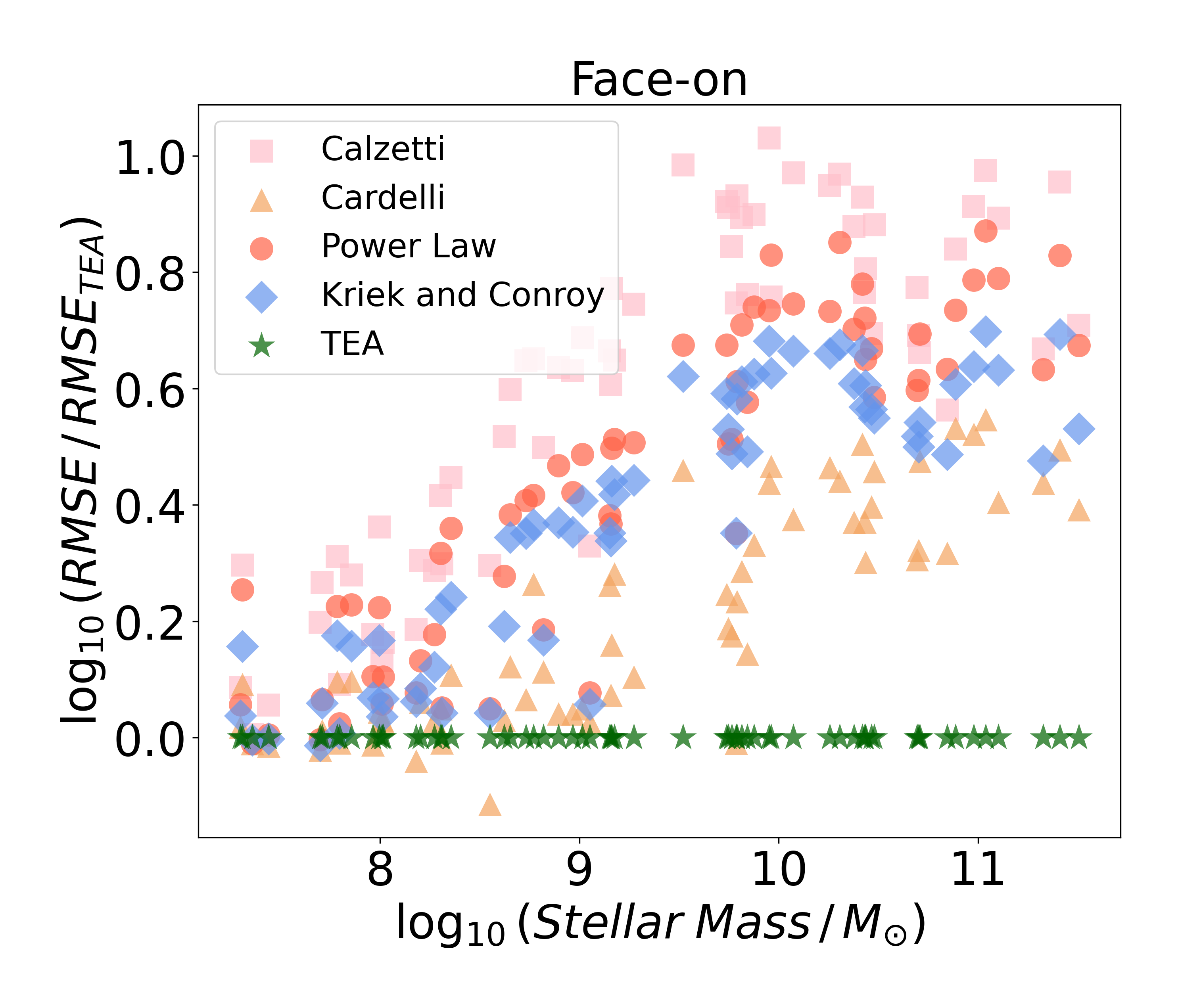}{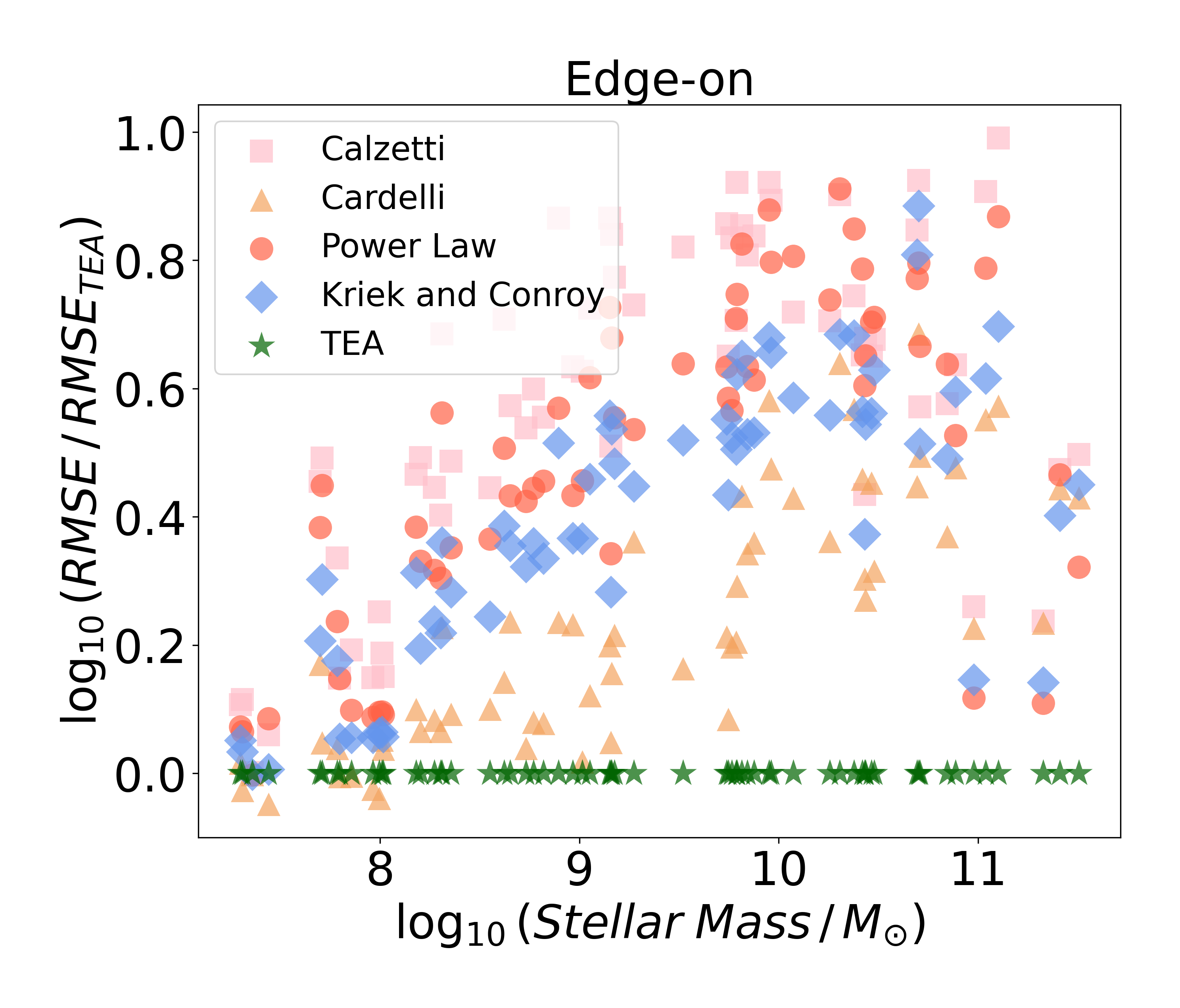}
\end{center}
\caption{\label{fig:shifted_errors2} \small
Similar to Figure \ref{fig:errors2}, now with the RMSEs shifted by the TEA model's RMSE for each galaxy.}
\end{figure}

Figure \ref{fig:errors2} shows the root mean squared error (RMSE) between the dust model fits and the simulated attenuation curves without allowing a fraction of starlight to reach the observer unattenuated. 
We can see that in this case, the TEA model outperforms nearly all other models for all galaxies and orientations. Figure \ref{fig:shifted_errors2} shows the RMSE, offset in log-space by the RMSE of the TEA model, as a function of galaxy stellar mass. This plot makes it easier to see how the different models' RMSEs compare to each other by subtracting away the scatter from the differences in RMSEs between galaxies.

\subsection{Fitting the TEA model to best-fitting common model attenuation curves} \label{Fitting the TEA model to best-fitting common model attenuation curves}

\noindent The TEA model is also flexible enough to reproduce the behavior of the 
other model attenuation curves. To demonstrate this fact we fit the TEA model to the other
models, using for the latter the
parameters from all of the best fits to the NIHAO-SKIRT-Catalog. 

\begin{figure}
\gridline{\fig{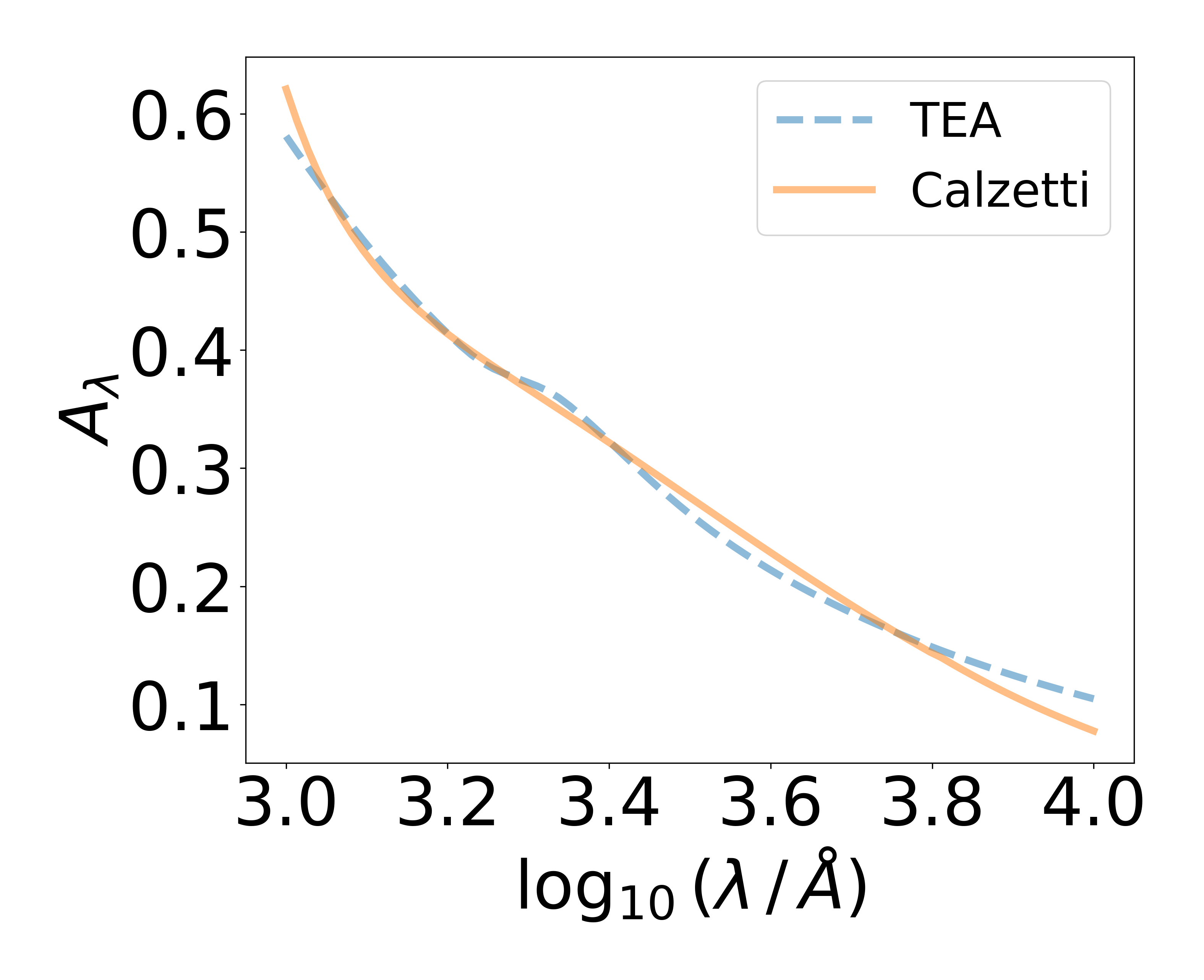}{0.3\textwidth}{(a) Calzetti}    \fig{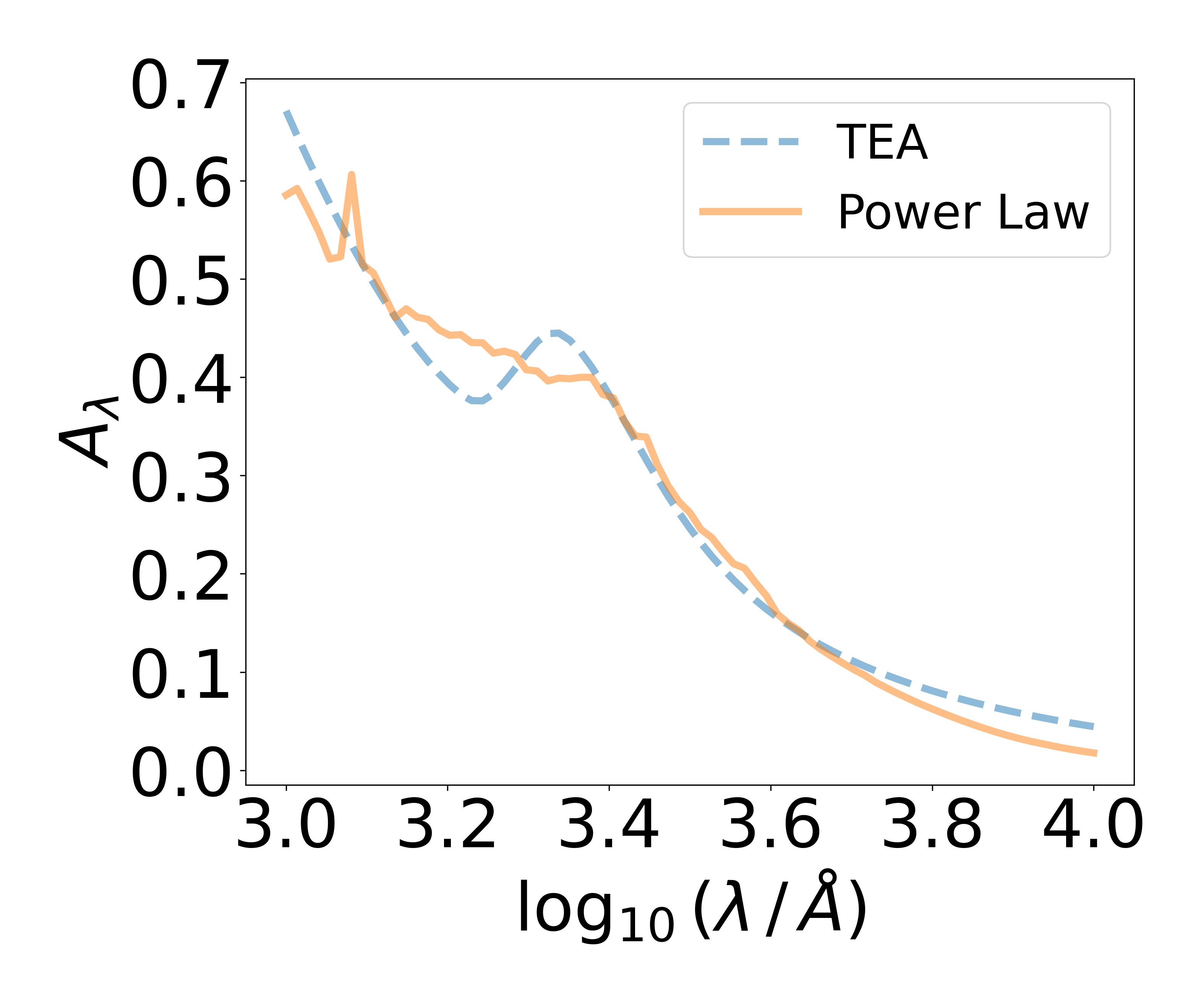}{0.3\textwidth}{(b) Power Law}}
\gridline{\fig{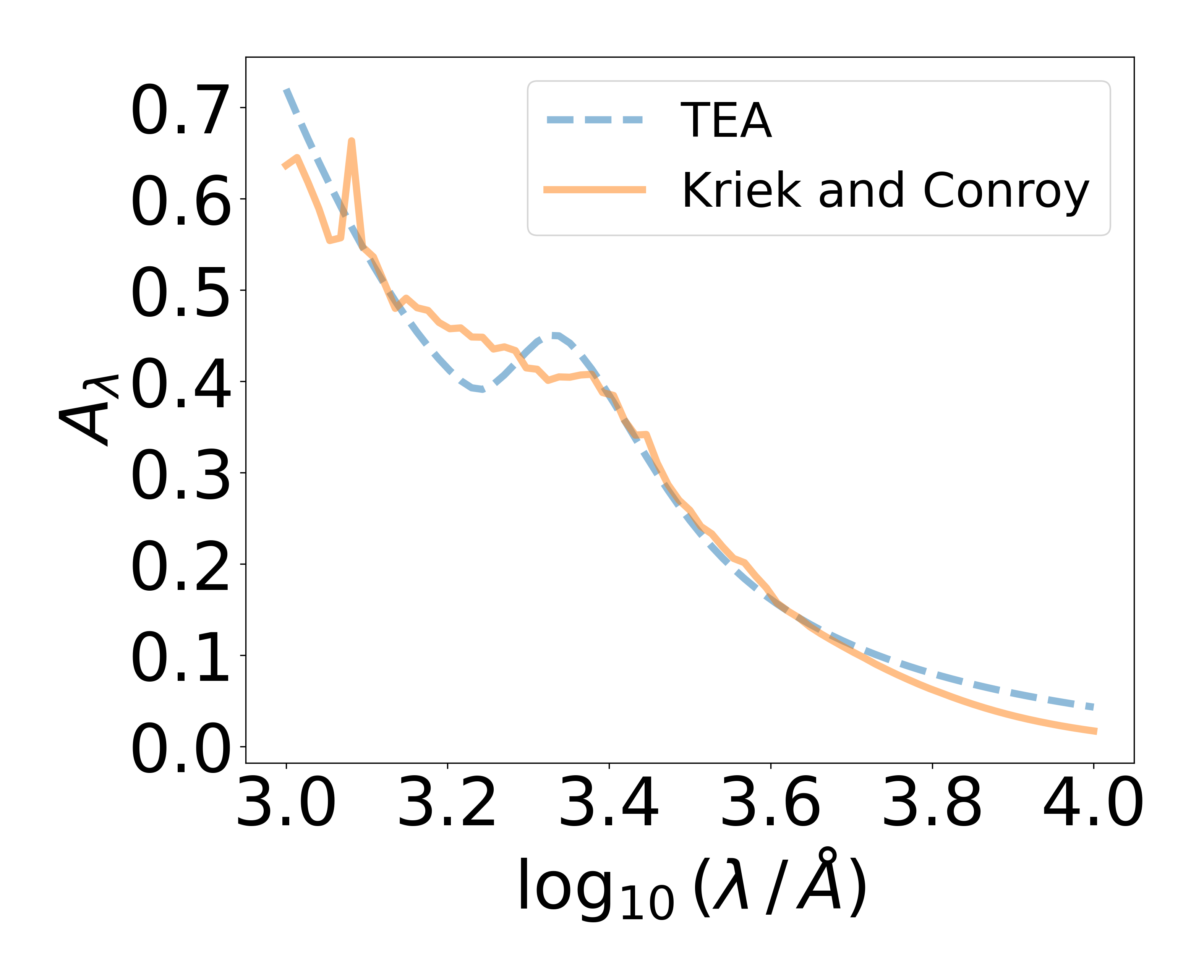}{0.3\textwidth}{(c) Kriek and Conroy}
\fig{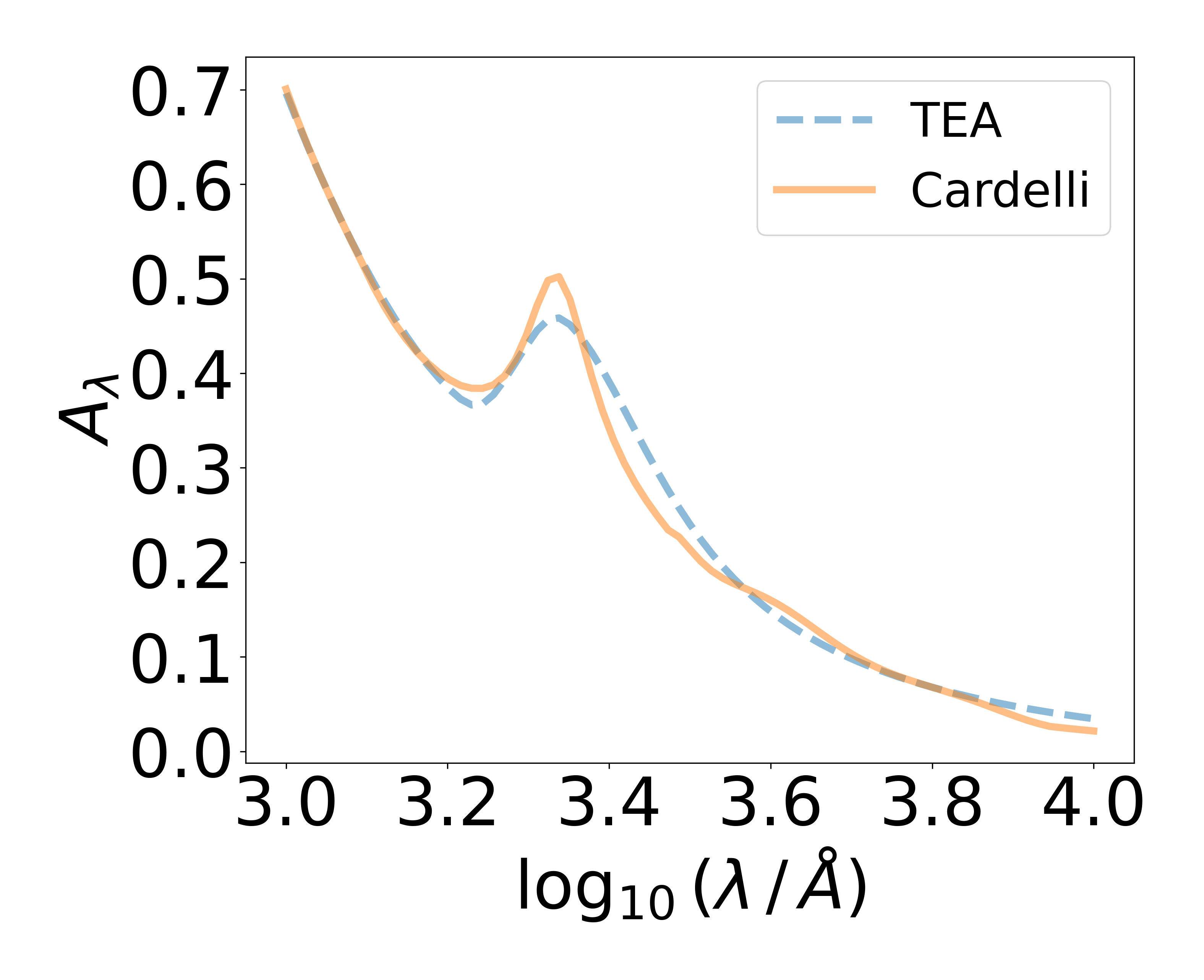}{0.3\textwidth}{(d) Cardelli}}
\caption{\label{fig:TEAToCommonFits}Fits of the TEA model to commonly used dust model fits of the face-on orientation of attenuation curve in magnitudes from the simulated galaxy {\tt g5.02e11}, allowing all parameters to be free in the fits. In this case, Calzetti has 2 free parameters, Power Law has 6 free parameters, Kriek and Conroy has 6 free parameters, Cardelli has 7 free parameters, and the TEA model has 3 free parameters.}
\end{figure}

Figure \ref{fig:TEAToCommonFits} shows the result of this process
applied to the face-on orientation of the simulated galaxy {\tt g5.02e11}, 
using the models that allow for freedom in 
$f_{\rm no-dust}$ and $f_{\rm no-dust, \, young}$. It may appear counter-intuitive at first glance that the TEA model fit to the power law curve has a strong UV-bump, since if we set $b_{\rm UV}$ to zero the TEA model gives exactly a power law. However, we must keep in mind that the power law model isn't a simple power law, its shape is complicated by its two-component nature as well as the flexibility to only apply attenuation to a fraction of the young and old stellar populations. Given that the TEA model has neither of these features, it uses the UV-bump to match the power law model's curve as well as possible. This same line of reasoning also applies to the TEA model fits to the Kriek and Conroy curves. 

\begin{figure}
\begin{center}
\plottwo{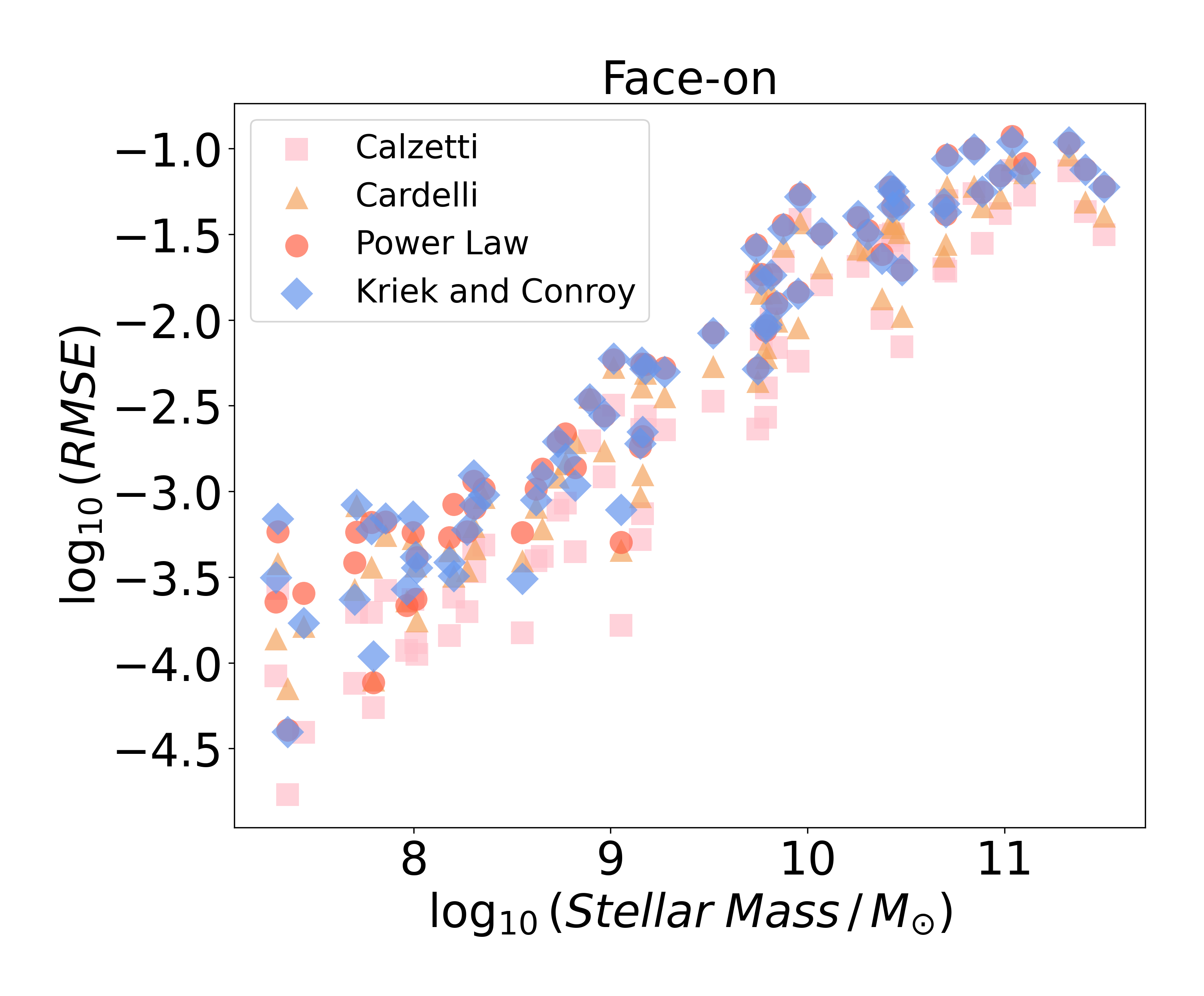}{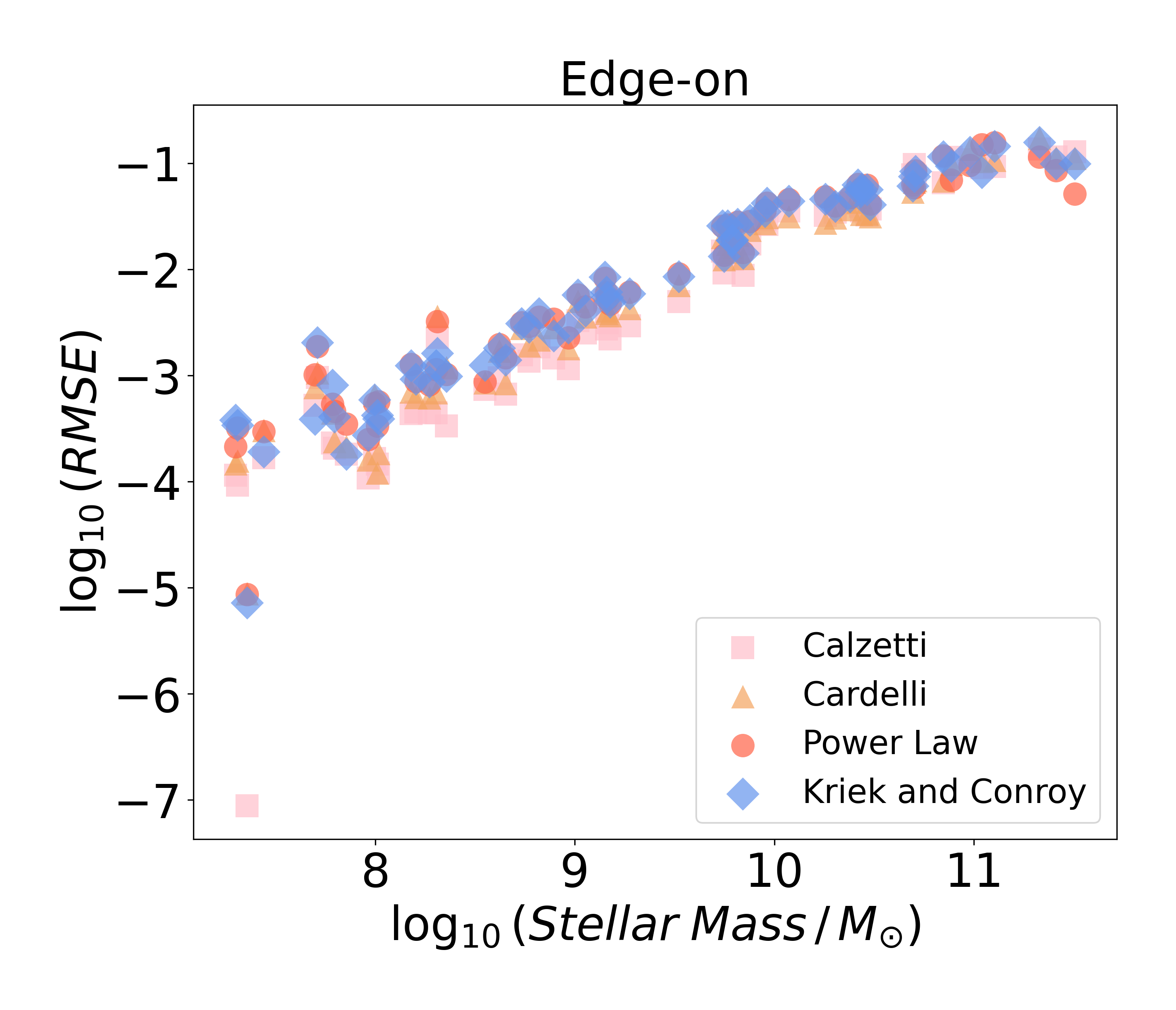}
\end{center}
\caption{\label{fig:TEAToCommonErrors} \small 
Root mean squared errors (RMSE) of the TEA model fit to commonly used model fits for the face-on (left) and edge-on (right) orientations of the simulated galaxies' attenuation curves in magnitudes.}
\end{figure}

Figure \ref{fig:TEAToCommonErrors} shows the root mean squared error (RMSE) between the TEA model fits and the best fits from the commonly used attenuation models of the simulated attenuation curves, allowing for a fraction of the stellar light to reach the observer unattenuated in the commonly used attenuation models. 

\begin{figure}
\gridline{\fig{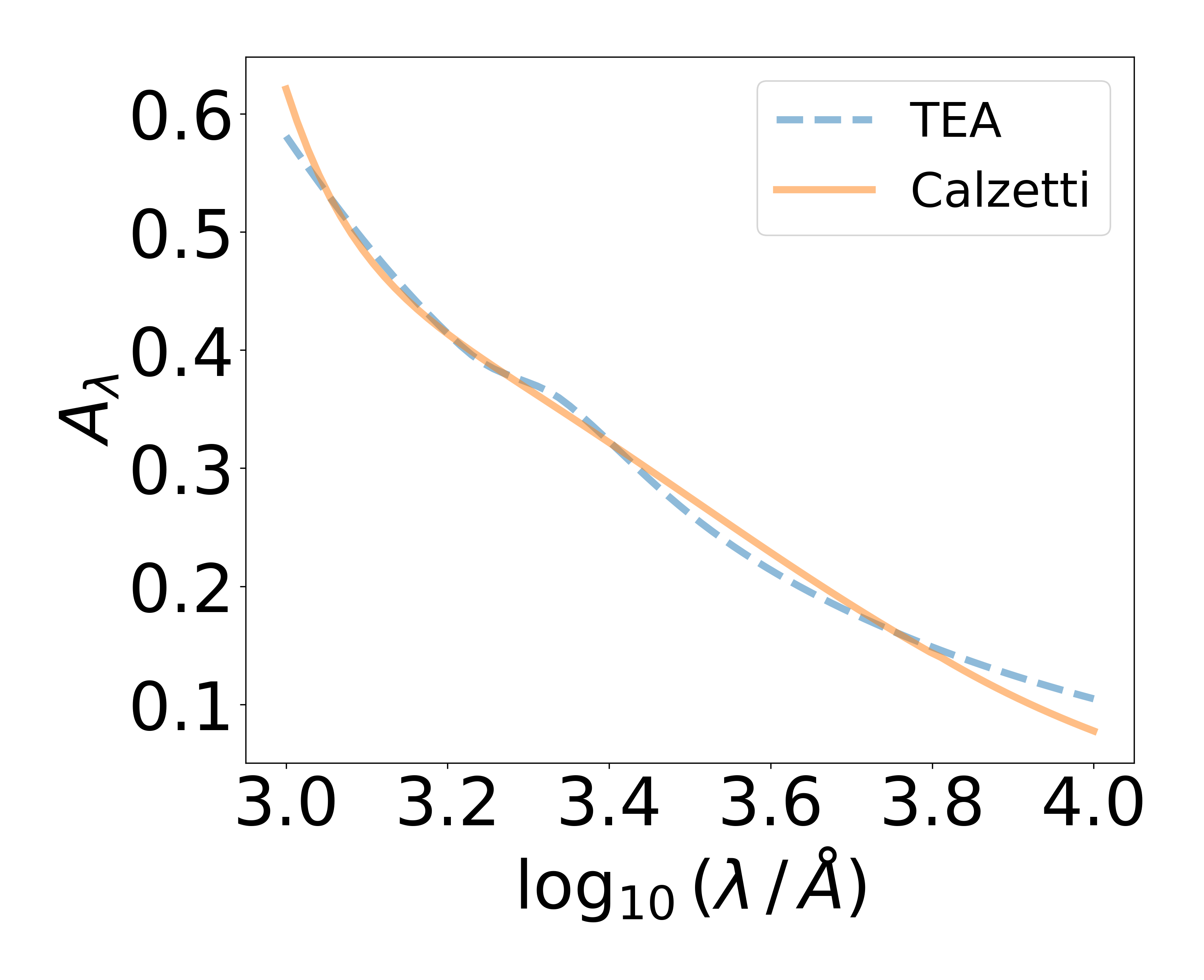}{0.3\textwidth}{(a) Calzetti}    \fig{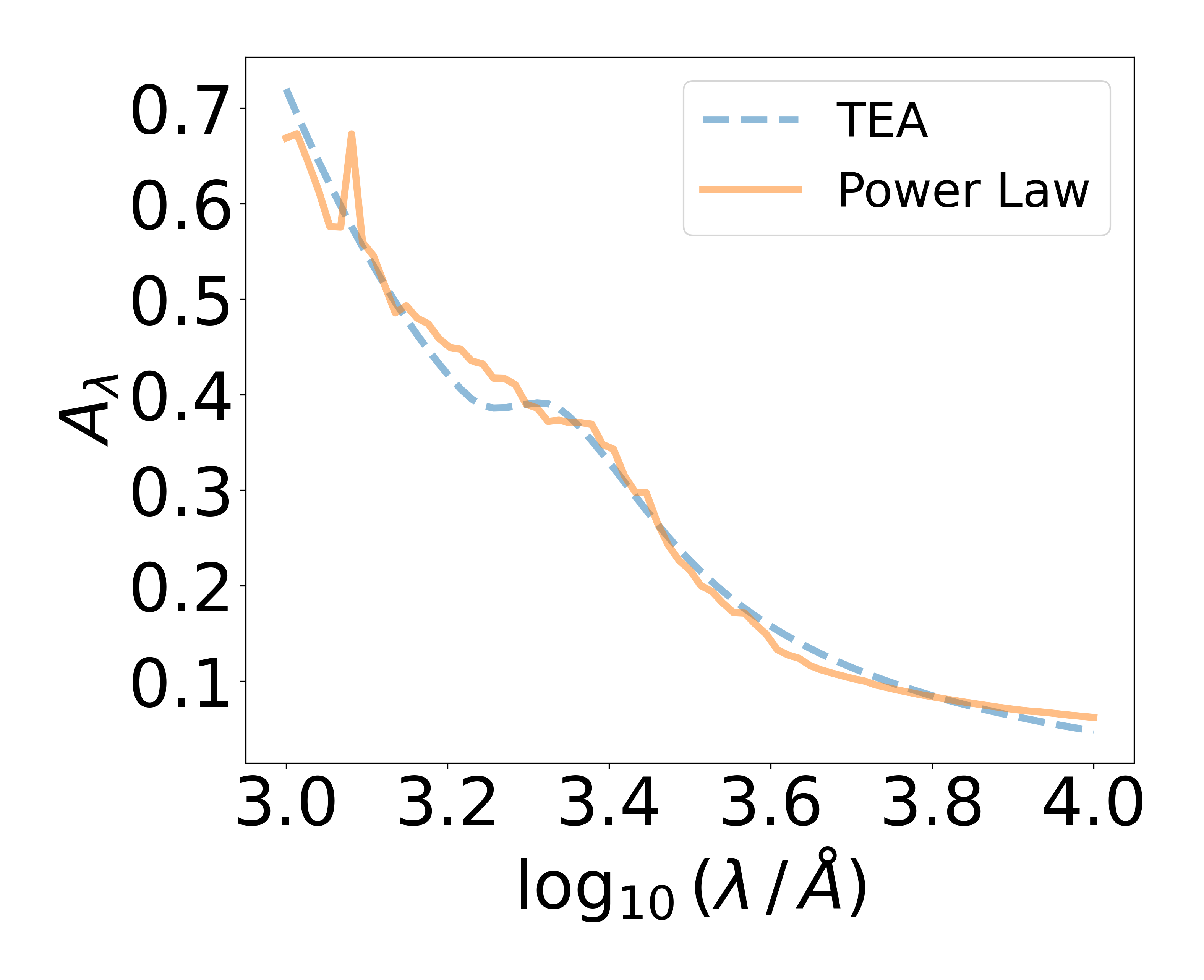}{0.3\textwidth}{(b) Power Law}}
\gridline{\fig{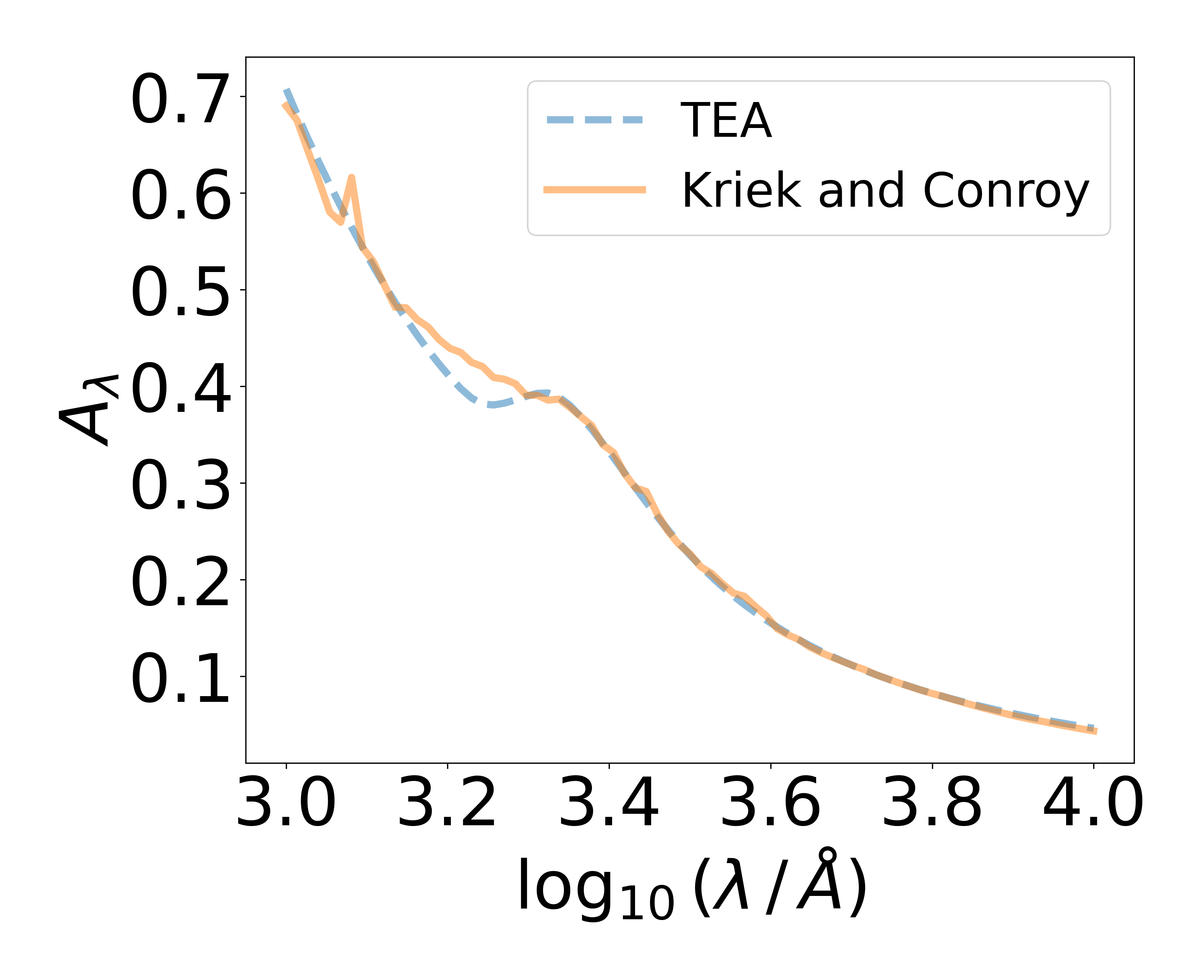}{0.3\textwidth}{(c) Kriek and Conroy}
\fig{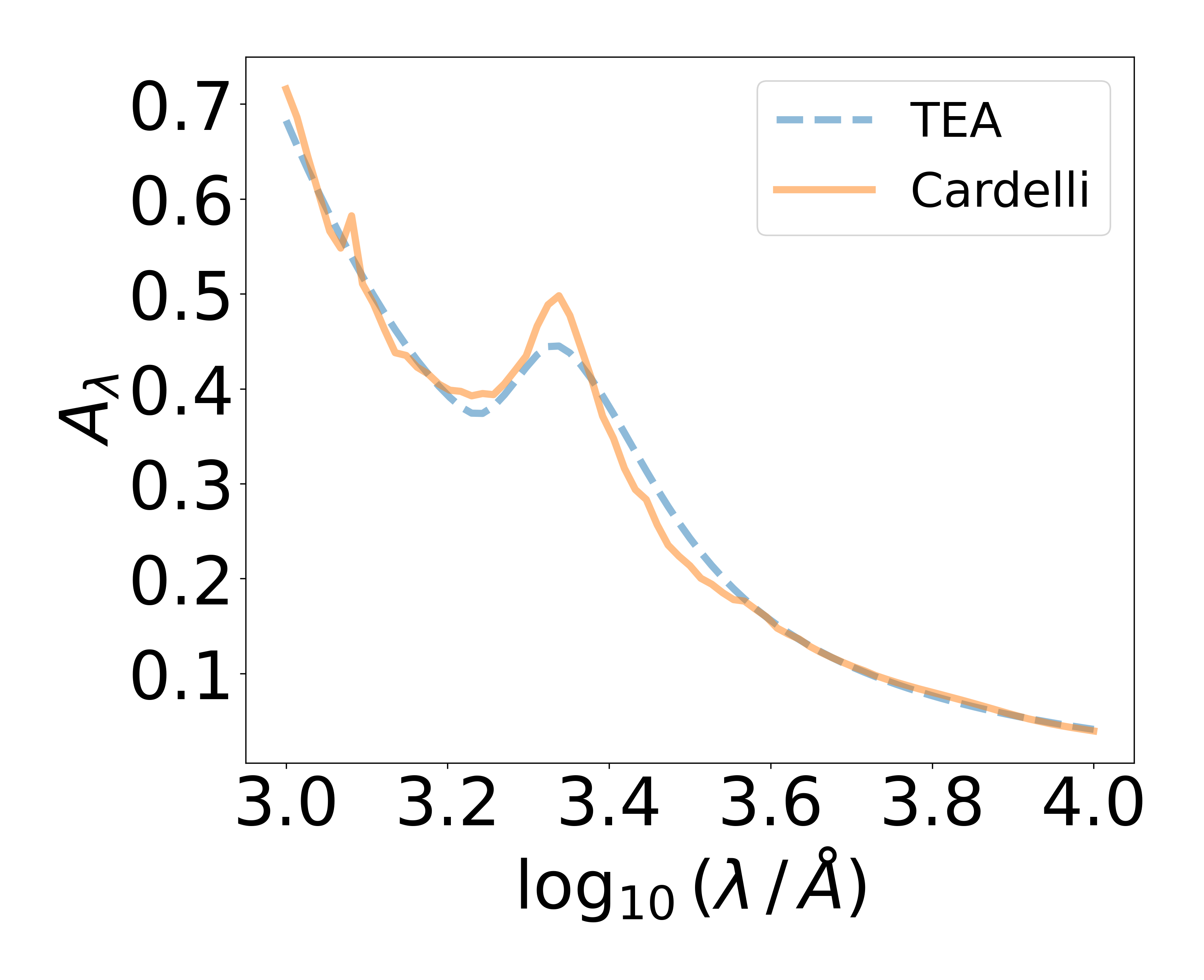}{0.3\textwidth}{(d) Cardelli}}
\caption{\label{fig:TEAToCommonFitsNoFrac}
Similar to Figure \ref{fig:TEAToCommonFits}, now not allowing any fraction of the stellar light to reach the observer unattenuated. In this case, Calzetti has 1 free parameter, Power Law has 4 free parameters, Kriek and Conroy has 4 free parameters, Cardelli has 5 free parameters, and TEA model has 3 free parameters.}
\end{figure}

Figure \ref{fig:TEAToCommonFitsNoFrac} shows the fits of the TEA model to the best fitting curves of the other dust models to the face-on orientation of the simulated galaxy {\tt g5.02e11}, 
fixing $f_{\rm no-dust} = f_{\rm no-dust, \, young}=0$, i.e. that all stellar light is attenuated in the commonly used attenuation models. Comparing the TEA model fit to the power law curves between Figures \ref{fig:TEAToCommonFits} and \ref{fig:TEAToCommonFitsNoFrac}, we can see that taking away the power law model's ability to only apply attenuation to a fraction of the young and old stellar light yields TEA model fits with weaker UV-bumps. We can understand this by recalling that the TEA model is using its UV-bump to compensate for its inability to separately attenuate the light from young and old stellar light (both figures), as well as its inability to only apply attenuation to a fraction of the young and old stellar light (Figure \ref{fig:TEAToCommonFits} only). 

\begin{figure}
\begin{center}
\plottwo{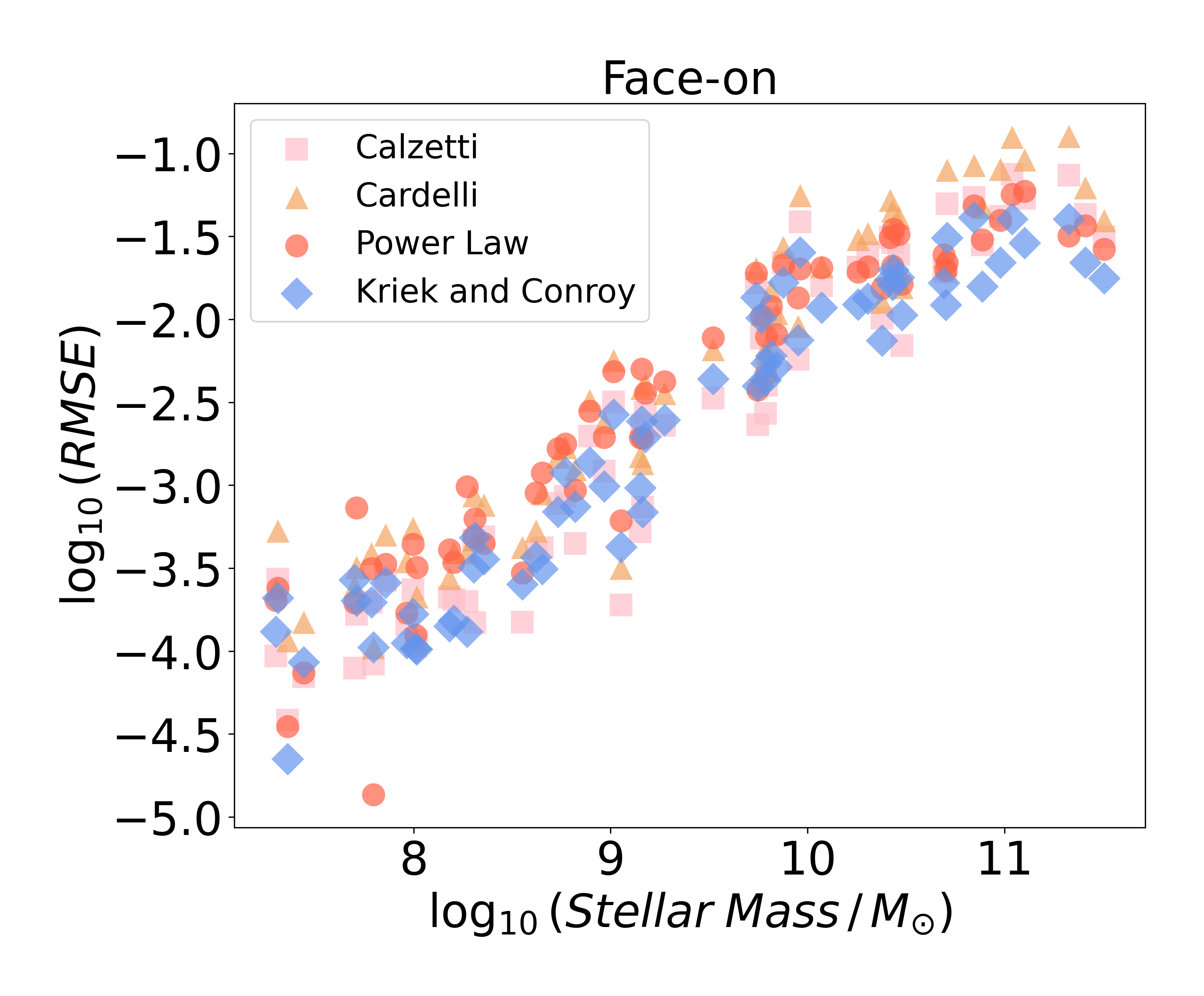}{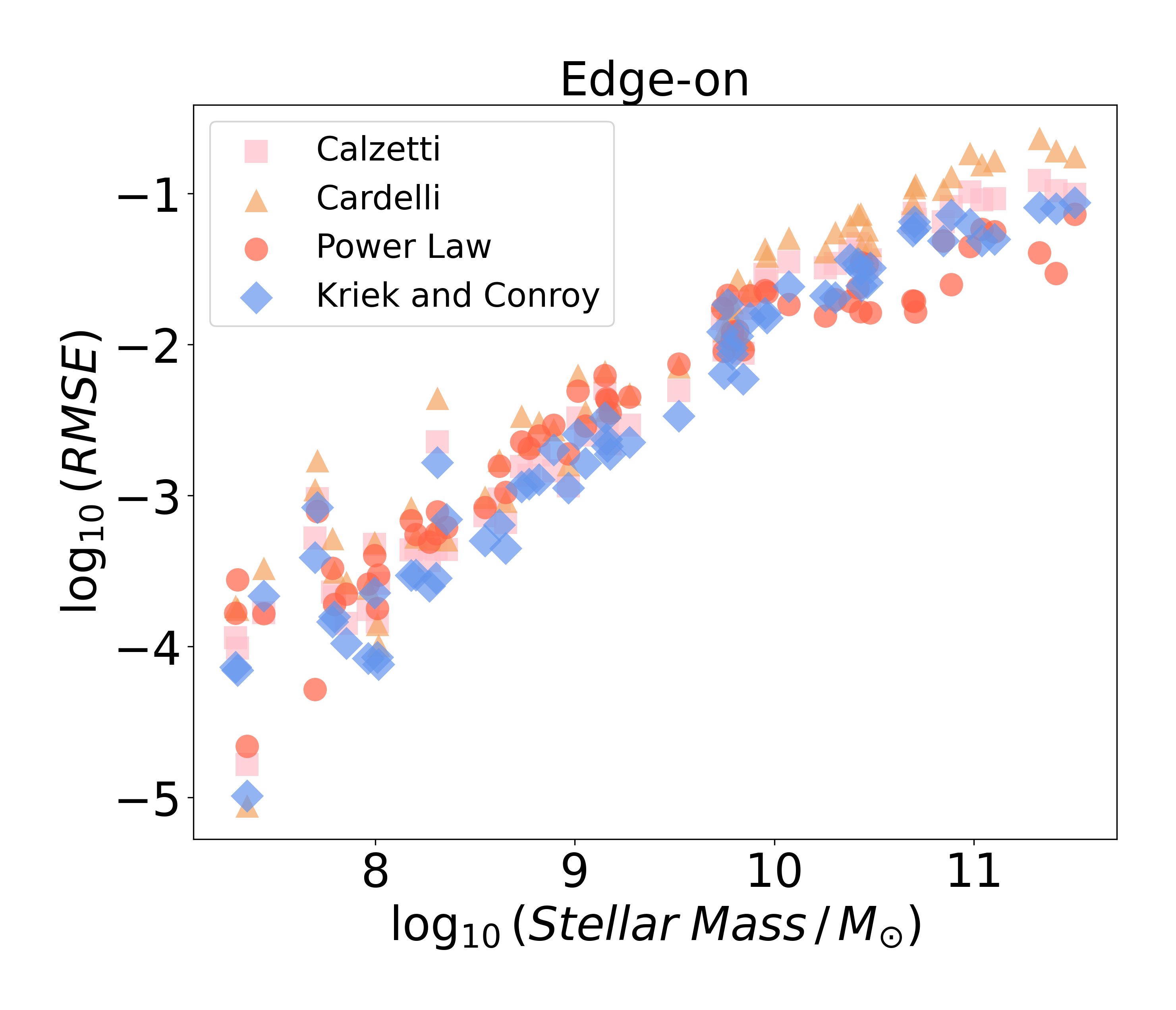}
\end{center}
\caption{\label{fig:TEAToCommonErrorsNoFrac} \small Similar to Figure \ref{fig:TEAToCommonErrors}, now not allowing any fraction of the stellar light to reach the observer unattenuated.}
\end{figure}

Figure \ref{fig:TEAToCommonErrorsNoFrac} shows the root mean squared error (RMSE) in magnitudes between the TEA model fits and the best fits from the commonly used attenuation models of the simulated attenuation curves, 
fixing $f_{\rm no-dust} = f_{\rm no-dust, \, young}=0$, i.e. that all stellar light is attenuated in the commonly used attenuation models. 

Comparing Figures \ref{fig:TEAToCommonErrors} and \ref{fig:TEAToCommonErrorsNoFrac} to Figures \ref{fig:errors1} and \ref{fig:errors2}, we can see that the TEA model is able to reproduce the best fitting attenuation curves of the common models better than the common models are able to reproduce the attenuation curves from the NIHAO-SKIRT-Catalog. 

\subsection{Correlation Between TEA model parameters} \label{Correlation Between TEA model parameters}

\begin{figure}
\begin{center}
\includegraphics[width=0.6\textwidth]{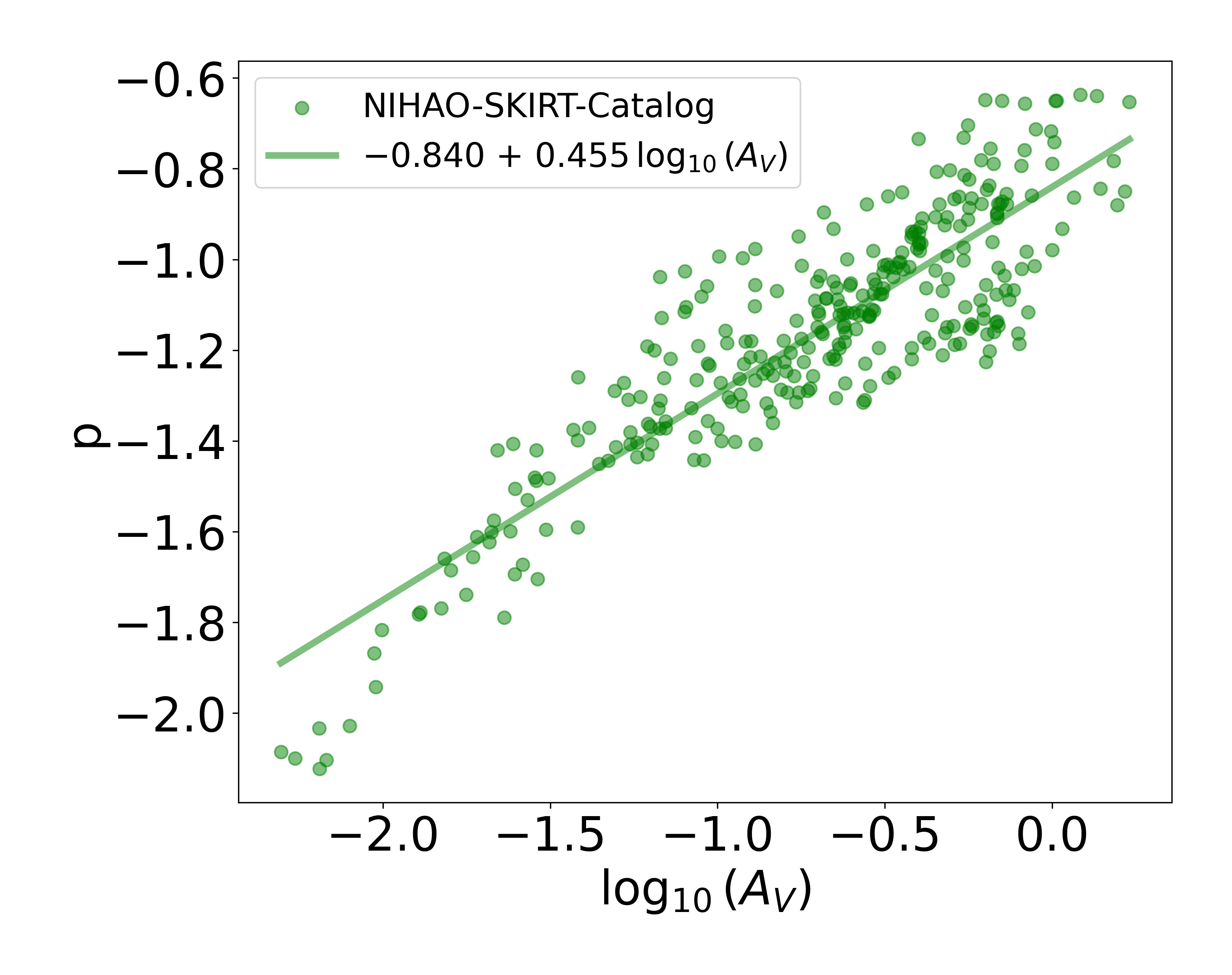}
\end{center}
\caption{\label{fig:Av_powerIndex} \small Distribution of best-fitting $p$ and $\log_{10}(A_{\rm V})$ parameters of the TEA model to the NIHAO-SKIRT-Catalog. Here we only include galaxies which have at least one orientation with an $A_{\rm V}$ value greater than 0.05 magnitudes. We also include the equation for the best-fitting line in the legend.}
\end{figure}

\noindent In this section we explore the TEA model's best-fitting parameters to the NIHAO-SKIRT-Catalog attenuation curves. We first consider the correlation between overall attenuation strength ($A_{\rm V}$) and the power law slope ($p$) shown in Figure \ref{fig:Av_powerIndex}. While there is a substantial amount of scatter in the relation, there is a clear trend for more heavily attenuated galaxies to have grayer (less steep) attenuation curves.

\begin{figure}
\begin{center}
\includegraphics[width=0.6\textwidth]{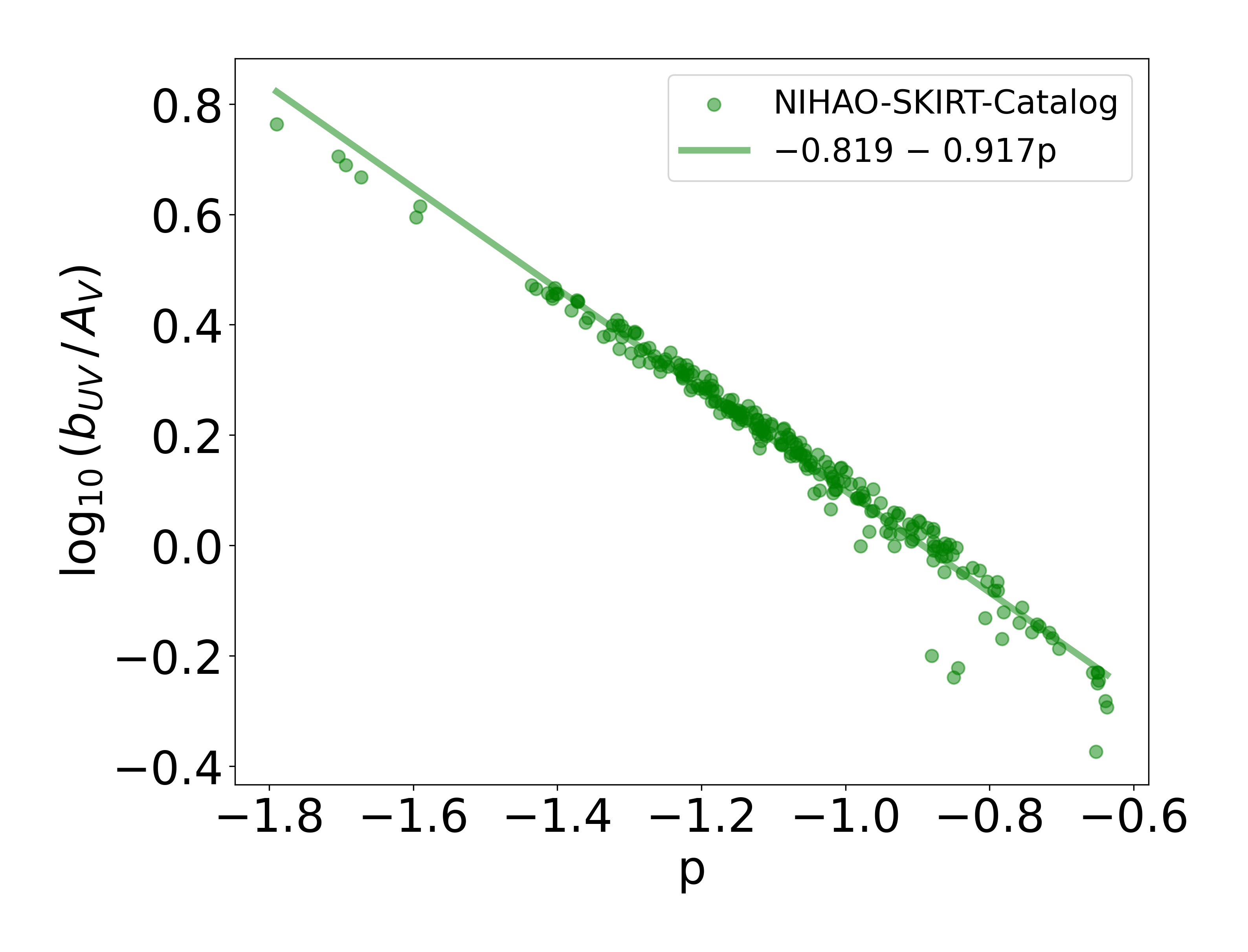}
\end{center}
\caption{\label{fig:powerIndex_bumpStrength} \small Distribution of best-fitting $\log_{10}(b_{\rm UV} \, / \, A_{\rm V})$ and $p$ parameters of the TEA model to the NIHAO-SKIRT-Catalog. Here we only include galaxies which have at least one orientation with an $A_{\rm V}$ value greater than 0.2. We also include the equation for the best-fitting line in the legend.}
\end{figure}

Motivated by the imposed relation between power law index and bump strength in the Kriek and Conroy model \citep{kriek13a}, we explore how power law index in the TEA model, $p$, relates to bump strength, $b_{\rm UV}$. Figure \ref{fig:powerIndex_bumpStrength} shows the distribution of best-fitting $\log_{10}(b_{\rm UV} \, / \, A_{\rm V})$ and $p$ parameters from the NIHAO-SKIRT-Catalog. To ensure we're limiting our analysis to sufficiently dusty galaxies, we only include galaxies which have at least one orientation with an $A_{\rm V}$ value greater than 0.2. We can see that the fits exhibit a strong correlation between these parameters, qualitatively consistent with the findings of \cite{kriek13a}. We note that this correlation is primarily due to the variation in the star-dust geometries of the simulated galaxies, since only one dust extinction curve is implemented in the radiative transfer calculations of the diffuse interstellar medium (ISM) in \cite{faucher23a}. 

To the extent that this correlation holds true for galaxies in the real universe, the best fitting line included in the legend of Figure \ref{fig:powerIndex_bumpStrength} may be used to reduce the dimensionality of the TEA model from 3 to 2 without significant loss of practical flexibility by parameterizing the bump strength in terms of $p$ and $A_{\rm V}$. Alternatively, one could rewrite the TEA model's bump term by replacing the free parameter $b_{\rm UV}$ with the offset from this relation, effectively reducing the size of the model's parameter space while retaining the flexibility to vary the bump strength and power law index independently. 

Due to the differences in the UV-bump shapes and functional forms of the Kriek and Conroy model (Equations \ref{eq:KC} and \ref{eq:KC_bump}) and the TEA model (Equation \ref{eq:tea}), it is difficult to directly compare the power law index to bump strength correlations between the two models. To circumvent this difficulty, we choose to explore this relation through the maximum UV-bump height relative to the overall attenuation, 
which we calculate by finding the maximum value of $A_{\lambda}/A_{\rm V}$ within the bump regions for the two models ($3.23 < \log_{10}(\lambda) < 3.6$ for the TEA model and $3.3 < \log_{10}(\lambda) < 3.4$ for the Kriek and Conroy model). 

As noted in Section \ref{Kriek and Conroy}, the effective power law index of the Kriek and Conroy model is offset from the value of its $p$ parameter by $\sim -0.8$ due to its Calzetti component. To account for this offset, we perform linear fits of the attenuation curves on a log-log scale, with their UV-bumps masked out, to find their effective power law indices. 

\begin{figure}
\begin{center}
\includegraphics[width=0.6\textwidth]{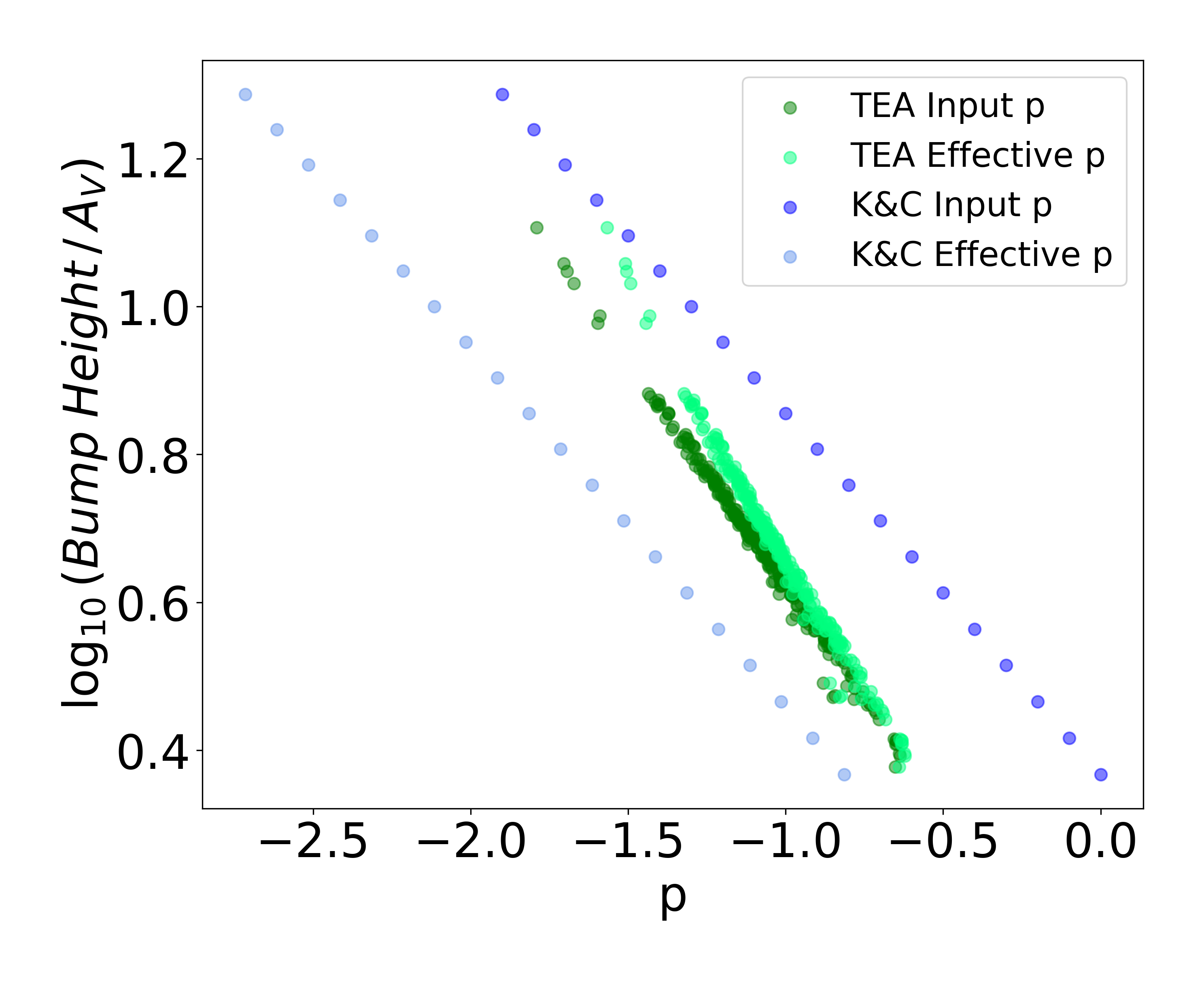}
\end{center}
\caption{\label{fig:powerIndex_bumpHeight} \small Comparison of the power law index to UV-bump height correlation between the TEA model and the Kriek and Conroy model. For both models, we calculate the effective power law index of the attenuation curves by performing a linear fit on log-log scale with the UV-bump regions masked out.}
\end{figure}

Figure \ref{fig:powerIndex_bumpHeight} shows how bump height correlates with the power law index model parameters and with 
the effective power law indices, for the TEA model fit to the NIHAO-SKIRT-Catalog galaxies and the Kriek and Conroy model, 
over a grid of power law index model parameters. 
Although there is some discrepancy in the slope and vertical offset of the relation, the TEA model fits follow a similar trend to the one imposed by the Kriek and Conroy model, with steeper curves correlating with stronger UV-bumps.

\section{Discussion} 
\label{Discussion}

\noindent Although the figures in the previous section demonstrate that the TEA model is 
able to accurately reproduce the attenuation curves from the NIHAO-SKIRT-Catalog, it is 
important to remember that the catalog attenuation curves are specific to the NIHAO 
simulations, the THEMIS dust model, and the sub-grid modeling of photo-disassociation 
regions (PDRs) around young stellar populations. 

In particular, all of these  attenuation curves include a bump around 2175 \AA, the UV-bump. 
A number of authors have concluded that the attenuation curves of some galaxies in 
the real universe such as the Small Magellanic Clouds (SMC) contain an extremely weak or absent UV-bump \citep{prevot84a, gordon98a,misselt99a, gordon03a}. Some of these conclusions
have been challenged; for instance \cite{calzetti94a} find that M33 
has no UV-bump but recent work by \cite{wang22a} 
suggests that it does. 

Regardless, we do know that, theoretically, different galaxies should exhibit 
different UV-bump strengths. We have designed the TEA model to have a variable 
UV-bump strength, enabling it to accurately reproduce attenuation curves with 
strong bumps, weak bumps, or no bump at all. This requirement was the primary 
motivation against using symbolic regression on the NIHAO-SKIRT-Catalog 
attenuation curves to find the optimal functional form, 
since based on that training data
that process would have tended to only select for strong UV-bumps. 

Although we have decided to use only a single component for the TEA model, 
this choice necessarily limits the resulting attenuation curves to be smooth 
and fully described by Equation \ref{eq:tea}. This contrasts to with two-component 
models, which can yield attenuation curves with small scale features due to the 
differences between the underlying young and old spectra. In particular, 
emission lines that tend to be stronger in young stellar populations will 
receive extra attenuation in two-component models that cannot be reproduced 
by single-component models. This limitation is unlikely to affect the flexibility 
when fitting only to broadband photometry, but two-component models may 
have an advantage when fitting moderate or high resolution spectra. Furthermore, 
despite being single-component, the TEA model is able to reproduce the attenuation 
curves of the NIHAO-SKIRT-Catalog, which does include additional extinction for 
young stellar populations, better than the existing two-component models 
explored in this work. This indicates that the flexibility to have small 
scale features in the attenuation curves of two-component models is less 
important that the large scale flexibility, at least relative to the loss function used in this work (the RMSE described in Section \ref{Fitting Procedure}). 

In addition to being able to accurately reproduce attenuation curves that 
account for complex star-dust geometries and birth cloud attenuation, 
the TEA model has the practical advantage of being fully analytic. This 
fact allows for convenient calculation of analytic gradients with respect to 
model parameters which can be used in optimization methods such as 
stochastic gradient descent (SGD). While this is also technically possible for the other dust models explored in this work, there are extra steps involved due to having two separate extinction curves for the young and old stellar populations (Power Law, Kriek and Conroy, and Cardelli), as well as having extinction curves with wavelength dependent functional forms (Calzetti, Kriek and Conroy, and Cardelli). As the need for computationally 
efficient SED modeling methods grows with the increase in astronomical 
data from large surveys, this ability to perform automatic differentiation 
will likely be extremely useful. 

The analysis performed in this work is not one that can be 
done for galaxies in the real universe, since attenuation curves are 
not a direct observable, but can only be inferred in the context of 
full SED modeling. In the context of SED modeling, attenuation curves are 
applied to stellar population spectra in order to match the models to the 
observed data, which could be in the form of broadband photometry 
and/or spectra. The parameter spaces of these models are often high-dimensional 
and lead to to loss-landscapes with many local minima, which provides 
further motivation for a dust attenuation model with as few 
free parameters as possible, but with sufficient flexibility. 

\section{Conclusion} 
\label{Conclusion}

\noindent In this work, we have tested the flexibility of commonly used 
dust attenuation models by using them to fit the attenuation curves 
from the NIHAO-SKIRT-Catalog. While the catalog attenuation curves are limited 
by the underlying simulations and dust models, they do include a physically 
motivated distribution of grain sizes and chemical compositions, nonlinear 
effects resulting from complex star-dust geometries, and sub-grid modeling 
of PDRs around young stellar populations. 

Motivated by the lack of flexibility of these commonly used attenuation 
models, we have proposed a new dust attenuation model that is able to 
accurately reproduce the catalog attenuation curves with only three free 
parameters. The new model, which we call the TEA model, is fully analytic 
and single-component, meaning its shape does not depend on the stellar 
spectra that it is applied to. This contrasts with two-component models 
that first attenuate the light coming from young stellar populations 
with a power law extinction curve before attenuating the composite spectra. 
The resulting shape of the global attenuation curve from two-component 
models will depend on the shape of the young stellar spectra relative 
to the old stellar spectra. These two characteristics make the TEA model 
both computationally efficient to model and easily compatible with automatic 
differentiation. 
In the era of big data in astronomy, these features make the 
TEA model particularly appealing in the context of SED modeling optimization. 

Using our parameterization, we have studied the relationship between 
the overall shape of the  attenuation curve and the strength of the 
2175 \AA\ bump. We find
that these two features to correlate well, in a sense that agrees 
qualitatively (though not quantitatively) with the work of \cite{kriek13a}---
grayer attenuation curves have smaller bump strengths.
Because the extinction law is constant within our models, we conclude 
that the physical cause of this close correlation is due to the 
star-dust geometry---patchier extinction leads to both grayer 
attenuation and smaller bump strengths.

We have made the TEA model, as well as Python versions of the commonly used 
dust attenuation models explored in this work, publicly available.\footnote{\url{https://github.com/ntf229/dustModels}}

\clearpage

\bibliography{refs}{}
\bibliographystyle{aasjournal}

\end{document}